\documentclass[%
reprint,
superscriptaddress,
nofootinbib,
amsmath,amssymb,
aps,
prd,
]{revtex4-2}
\usepackage[dvipdfmx]{graphicx}
\usepackage{dcolumn}
\usepackage{bm}
\usepackage{orcidlink}

\usepackage{physics}

\usepackage[normalem]{ulem} 

\usepackage{aas_macros}

\usepackage{subfigure}

\usepackage{booktabs} 
\usepackage{siunitx}  
\usepackage{caption} 
\usepackage{enumerate}

\usepackage{tikz}
\usepackage{xcolor}
\definecolor{RED}{rgb}{1,0,0} 

\usepackage{ragged2e} 

\usepackage{ulem} 

\usepackage[utf8]{inputenc}

\usepackage{multirow} 

\usepackage{float} 
\usepackage{placeins}

\usepackage{hyperref}

\usepackage[all]{hypcap}

\begin{document}

\preprint{APS/123-QED}

\title{One-dimensional PIC Simulation of Induced Compton Scattering in Magnetized Electron--Positron Pair Plasma
}

\author{Shoma F. Kamijima\orcidlink{0000-0002-4821-170X}}%
\email{shoma.kamijima@yukawa.kyoto-u.ac.jp}
\affiliation{%
 Center for Gravitational Physics and Quantum Information, 
 Yukawa Institute for Theoretical Physics, Kyoto University, Kyoto 606-8502, Japan}%
\author{Rei Nishiura\orcidlink{0009-0003-8209-5030}}
 \email{nishiura@tap.scphys.kyoto-u.ac.jp}
 \affiliation{%
 Department of Physics, Kyoto University, Kyoto 606-8502, Japan}%
\author{Masanori Iwamoto\orcidlink{0000-0003-2255-5229}}%
 \email{m-iwamoto@people.kobe-u.ac.jp}
\affiliation{%
 Graduate School of System Informatics, Kobe University, 1-1 Rokkodai-cho, Nada, Kobe, Hyogo 657-8501, Japan}%
\author{Kunihito Ioka\orcidlink{0000-0002-3517-1956}}%
 \email{kunihito.ioka@yukawa.kyoto-u.ac.jp}
\affiliation{%
 Center for Gravitational Physics and Quantum Information, 
 Yukawa Institute for Theoretical Physics, Kyoto University, Kyoto 606-8502, Japan}%



\date{\today}

\begin{abstract}
We investigate induced Compton scattering of a circularly polarized Alfv\'en wave propagating in a magnetized electron-positron pair plasma using one-dimensional Particle-in-Cell (PIC) simulations. In this system, two distinct modes of density fluctuations, referred to as the charged mode and the neutral mode, are theoretically expected to arise through parametric instabilities. Our simulations confirm these predictions: in the charged mode, the electron and positron densities fluctuate oppositely (Langmuir-like), while in the neutral mode, the charge is Debye-screened and both species fluctuate in phase (acoustic-like). The linear growth rates obtained from the simulations are in good agreement with analytical estimates for both modes. We also find that, in some cases, the linear growth saturates before full scattering occurs, allowing the incident wave to propagate without significant attenuation. Our results allow us to determine whether induced Compton scattering grows linearly in magnetized pair plasmas, offering a foundation for studies of fast radio bursts and laser-plasma experiments.
\end{abstract}

\maketitle


\section{Introduction}
Nonlinear interactions between plasmas and electromagnetic waves have been actively studied in both astrophysics and laboratory settings \cite{kaw73,max74,forslund75,tabak94,deutsch96,kwan79,friedland80}. 
The nonlinear interactions
can give rise to a wide range of plasma instabilities, including stimulated/induced Raman scattering, stimulated/induced Brillouin scattering, stimulated/induced Compton scattering, two-plasmon decay instability, oscillating two-stream instability, filamentation instability, and modulation instability \cite{sagdeev69,max73,max74,drake74,forslund75,mima75,cohen79,kruer88}.
Such nonlinear wave-plasma interactions are of interest in many kinds of astrophysical environments
\cite{galeev63,barnes66,derby78,goldstein78,terasawa86,longtin86,hoshino89,inhester90,vinas91,jayanti93,hollweg94,delzanna01,suzuki05,suzuki06,nariyuki06,nariyuki08,delzenna15,shoda16,shi17,shoda18,nariyuki22,thide82,robinson89,syunyaev71,coppi93,max73,blandford76,wilson78,wilson82,lyubarskii96,lyubarsky08,iwamoto23,sobacchi23,ishizaki24,sobacchi24}.

Recently, the importance of nonlinear plasma interactions has been highlighted in the context of fast radio bursts (FRBs) \cite{ghosh22,golbraikh23,sobacchi23,ishizaki24,nishiura25a,komissarov25,nishiura25b}. 
FRBs are the brightest radio transients in the Universe and are coherent emissions in the sub-GHz--GHz band with millisecond durations, first discovered in 2007 \cite{lorimer07}.
However, their emission mechanism and origins remain unclear \cite{petroff19,lyubarsky21,bzhang23}. 
While most FRBs are of extragalactic origin \cite{thornton13,cordes16,chime21}, the detection of FRBs from a Galactic magnetar in 2020 established that at least one FRB originates from a magnetar \cite{mereghetti20,bochenek20,chime20,snzhang20,li21,ridnaia21}. 
Magnetar models for FRBs are classified into two categories: the magnetosphere model  \cite{katz14,bzhang17,kumar17,ghisellini18,lu18,kumar20a,kumar20b,lu20,yuan20,ioka20,cooper21,qu24} and the wind model \cite{lyubarsky14,waxman17,beloborodov17,metzger19,beloborodov20,margalit20a,margalit20b,iwamoto24,vanthieghem25a}.
Although it is debatable which model is more plausible, both scenarios involve electromagnetic waves propagating through a magnetized electron-positron plasma.

\begin{figure*}[htbp]
    \centering
    \includegraphics[width=0.94\linewidth]{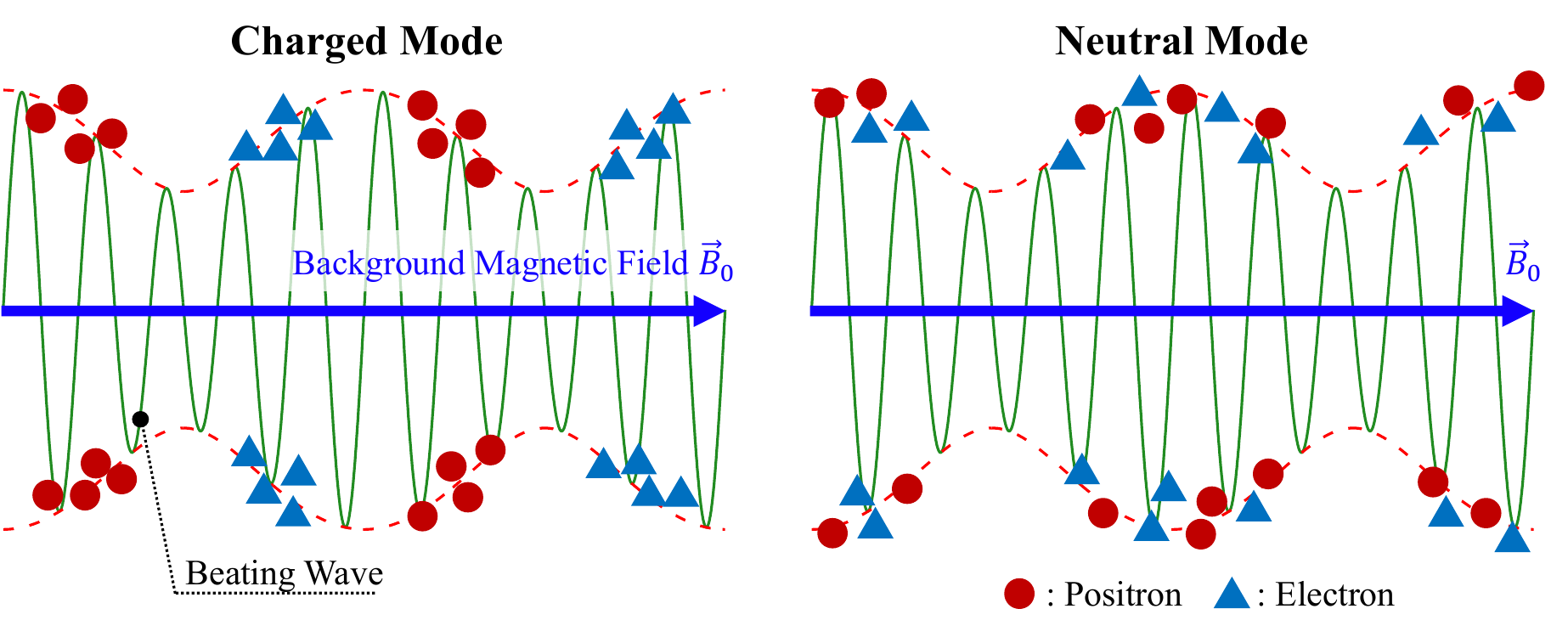}
    \caption{\RaggedRight 
    Schematic of the density fluctuation for the charged mode (left panel) and the neutral mode (right panel).
    Red circles and blue triangles means positrons and electrons.
    The green curve is the beating wave between the incident and scattered wave.
    The blue arrow is the background magnetic field. 
    For the charged (neutral) mode , electrons and positrons accumulate in the different (same) region due to the ponderomotive force (see second and third terms of Eq.~(14) in Ref.~\cite{nishiura25a})
    }
    \label{fig:mode}
\end{figure*}
This study mainly focuses on a magnetar magnetosphere model. 
One of the magnetar magnetosphere models suggests that Alfv\'en waves generated by starquakes or magnetic reconnection near the magnetar transport energy outward and produce the FRB emission via some mechanisms \cite{kumar17,lu18,kumar20a,kumar20b,lu20,yuan20,ioka20,qu24}. 
In the electron-positron pair plasma of a magnetar magnetosphere, induced Compton scattering has been suggested to prevent propagation of electromagnetic waves \cite{lyubarsky08,lyubarsky21,ghosh22,nishiura24,nishiura25a,nishiura25b}. 
In the classical interpretation of induced Compton scattering, an incident (parent, pump) wave propagates through the plasma, generating a scattered (daughter) wave propagating in the opposite direction, along with density fluctuations arising from the beating between the incident and scattered waves through parametric instability \cite{drake74}. 
Kinetic effects, particularly the (nonlinear) Landau resonance between the beating wave and plasma particles, play a crucial role in induced Compton scattering, mediating energy transfer from the wave to the particles.

The background magnetic field can affect on the growth rate of the induced Compton scattering.
For the case that the electric field is parallel to the background magnetic field (e.g. O-mode waves), the linear growth rate of the induced Compton scattering in the magnetized electron-positron pair plasma is same as that in the unmagnetized electron-positron pair plasma \cite{nishiura25a,nishiura25b,ghosh22}.
For the case that the electric field is perpendicular to the background magnetic field (e.g. X-mode waves and Alfv\'en waves), the linear growth rate of the induced Compton scattering in the magnetized electron-positron pair plasma can be reduced, compared to that in the unmagnetized electron-positron pair plasma \cite{nishiura25a,nishiura25b}.
This is because particle motion is constrained in the presence of a background magnetic field, which limits the plasma response to electromagnetic waves \cite{nishiura24,nishiura25a,nishiura25b}.
In addition to the effect of the background magnetic field, our previous studies show that the Debye screening can suppress the linear growth rate of induced Compton scattering in a magnetized electron-positron pair plasma \cite{nishiura25a,nishiura25b}.

Our previous studies analytically derived the linear growth rate of the induced Compton scattering in the magnetized electron-positron pair plasma and have indicated that, in a magnetized electron-positron pair plasma, two distinct modes—referred to as the charged mode and the neutral mode—can arise for the case that the electric field is perpendicular to the background magnetic field (e.g. X-mode waves and Alfv\'en waves) \cite{nishiura25a,nishiura25b}.
Different modes are generated by distinct mechanisms of density fluctuations, which arise from the ponderomotive force.
For X-mode and Alfv\'en waves in magnetized plasma, the ponderomotive potential consists of two contributions \cite{lee83,lee96,klima66,klima68,hatori81,cary77,cary81}:
\begin{enumerate}
\item a charge-sign-dependent term (see third term of Eq.~(14) in Ref.~\cite{nishiura25a}), which governs the charged mode,
\item a charge-sign-independent term (see second term of Eq.~(14) in Ref.~\cite{nishiura25a}), which governs the neutral mode.
\end{enumerate}
Figure \ref{fig:mode} shows the schematic of the density fluctuation for the charged mode (left) and neutral mode (right).
Red circles and blue triangles mean positrons and electrons, respectively.
The green curve and blue arrow are the beating wave between the incident and scattered waves and the background magnetic field, respectively.
For the charged (neutral) mode, electrons and positrons concentrate on the different (same) region owing to the related ponderomotive potential.
Although several particle-in-cell (PIC) simulations have been performed for electromagnetic waves or Alfv\'en waves propagating in a magnetized electron-positron pair plasma \cite{matsukiyo03,matsukiyo09,lopez14,munoz14}, it remains unclear whether both the charged and neutral modes can be realized and whether the analytical linear growth rate of both charged and neutral modes \cite{nishiura25a,nishiura25b} can be obtained from kinetic plasma simulations.

In this study, we focus on the propagation of Alfv\'en waves in an electron-positron pair plasma under a strong background magnetic field. 
We investigate induced Compton scattering of circularly polarized Alfv\'en waves propagating in a magnetized electron-positron pair plasma by using PIC simulations, examining whether the charged and neutral modes are realized and whether the linear growth rates are consistent with analytical estimates. 
In Sec.~\ref{sec:analytical}, we present the linear growth rates of the charged and neutral modes for circularly polarized waves. 
Section~\ref{sec:setup} describes the setup of the PIC simulations. 
Section~\ref{sec:result} presents the simulation results and saturation of induced Compton scattering.
Section~\ref{sec:summary} summarizes this study, and discusses its implications for FRBs.
Throughout this paper, italic symbols \(e\) denote the elementary charge, whereas roman type \(e\) represents the exponential \( \rm e = \exp(1) \).

\section{Linear Growth Rate of Induced Compton Scattering in Strongly Magnetized electron-positron pair plasma} \label{sec:analytical}
In this section, we estimate the linear growth rates of the charged and neutral modes of induced Compton scattering in a strongly magnetized electron-positron pair plasma for circularly polarized waves. 
In our previous paper \cite{nishiura25a}, the formulation can be applied regardless of polarization, whether linear or circular; however, the final expression assumes that the incident wave is linearly polarized. 
Here, we rewrite the expression for the linear growth rate for the case of circularly polarized waves used in our simulations.
We begin with the charged mode of induced Compton scattering \cite{nishiura25a}. 
For a linearly polarized incident wave, the maximum linear growth rate of the scattered wave energy, $\Gamma_{\rm C,max}^{\rm charged}$, is given in Eq.~(72) of Ref.~\cite{nishiura25a} as
\begin{eqnarray}
    \Gamma_{\rm C,max}^{\rm charged} &=& 2~{\rm Im}(\omega_1) , \nonumber \\
    &=&
    \sqrt{\frac{32 \rm e}{\pi}} \frac{k_{\rm B}T_{\rm e}}{m_{\rm e} c^2}
    \left( \frac{\omega_0}{\omega_{\rm c}} \right)^2
    \left( 1 + \frac{\omega_{\rm p}^2}{\omega_{\rm c}^2} \right)\nonumber  \left( \frac{\omega_0}{\omega_{\rm p}} \right)^4
    \frac{a_{\rm e}^2 \omega_{\rm p}^2}{\omega_0}, \nonumber \\
\end{eqnarray}
where $\omega_0$ and $\omega_1$ are the angular frequencies of the forward-propagating incident  wave and the backward-propagating scattered wave, respectively. 
Here, $k_{\rm B}$ is the Boltzmann constant, and $T_{\rm e}$ is the electron temperature, which is assumed to be equal to the positron temperature. 
The cyclotron frequency is defined as 
\begin{eqnarray}
    \omega_{\rm c} = \frac{e B_0}{m_{\rm e} c} \label{eq:wc}, 
\end{eqnarray} 
where $m_{\rm e}$ denotes the electron mass, $c$ is the speed of light, and $B_0$ is the background magnetic field strength. 
The plasma frequency $\omega_{\rm p}$ is given by
\begin{eqnarray}
    \omega_{\rm p} = \sqrt{2}\,\omega_{\rm pe} = \sqrt{ \frac{8\pi n_0 e^2}{m_{\rm e}} } \label{eq:wp},
\end{eqnarray}
where $\omega_{\rm pe}$ is the electron plasma frequency.
$n_0$ is the unperturbed density of the electron and positron plasma, which is assumed to be homogeneous and neutral.
In this study, we consider the strong background magnetic field.
Hence, the magnetization parameter $\sigma$ is assumed to be much larger than unity:
\begin{eqnarray}
    \sigma = \frac{\sigma_{\rm e}}{2}=\frac{B_0^2}{8\pi n_0 m_{\rm e}c^2} \gg 1 \label{eq:sigma},
\end{eqnarray}
where $\sigma_{\rm e}$ is the magnetization parameter for electrons.
From Eqs.~(\ref{eq:wc}),(\ref{eq:wp}), and (\ref{eq:sigma}), the cyclotron frequency is assumed to be much larger than the plasma frequency as follows:
\begin{eqnarray}
    \omega_{\rm c} = \sqrt{\sigma}\omega_{\rm p} \gg \omega_{\rm p} \label{eq:wcwp}. 
\end{eqnarray} 
The strength parameter for a linearly polarized incident wave $a_{\rm e}$ is given by
\begin{eqnarray}
    a_{\rm e} = \frac{2eA_0}{m_{\rm e}c^2},
    \label{eq:ae_lin} 
\end{eqnarray}
where $2A_0$ is the real magnitude of the vector potential of the incident wave (see Eqs.~(1) and (26) in Ref.~\cite{nishiura25b}).

In this study, we consider a circularly polarized incident wave.  
Even in this case, the growth rate has the same form when expressed in terms of $A_0$; however, since the strength parameter for a circularly polarized wave is defined differently in terms of $A_0$ (see Eq.~(A2) in Ref.~\cite{nishiura25b}) 
\begin{eqnarray}
    a_{\rm e}^{\rm circ} = \frac{\sqrt{2}eA_0}{m_{\rm e}c^2}.
    \label{eq:ae_circ}
\end{eqnarray}
Then, it is necessary to make the following replacement to express the growth rate in terms of $a_e^{\rm circ}$ (see Eq.~(\ref{eq:ae_circ}))
\begin{eqnarray}
    a_{\rm e} \rightarrow \sqrt{2} a_{\rm e}^{\rm circ}.
    \label{eq:ae_rel}
\end{eqnarray}
Therefore, using Eq.~(\ref{eq:ae_rel}),
the maximum growth rate for a circularly polarized incident wave $\Gamma_{\rm C,max}^{\rm circ,charged}$  
can be expressed in terms of the circularly polarized strength parameter $a_{\rm e}^{\rm circ}$ as follows:  
\begin{eqnarray}
    \Gamma_{\rm C,max}^{\rm circ,charged} &\simeq& 
    \sqrt{\frac{128 \rm e}{\pi}} \frac{k_{\rm B}T_{\rm e}}{m_{\rm e} c^2}
    \left( \frac{\omega_0}{\omega_{\rm c}} \right)^2
    \left( 1 + \frac{\omega_{\rm p}^2}{\omega_{\rm c}^2} \right)\nonumber \\
    &&\times
    \left( \frac{\omega_0}{\omega_{\rm p}} \right)^4
    \frac{\left(a_{\rm e}^{\rm circ}\right)^2 \omega_{\rm p}^2}{\omega_0} . \label{eq:gr_circ_c}
\end{eqnarray}

Next, we consider the neutral mode.  
For a linearly polarized incident wave, the maximum growth rate of the scattered wave energy
$\Gamma_{\rm C,max}^{\rm neutral}$ is given by Eq.~(96) in Ref.~\cite{nishiura25a} as
\begin{eqnarray}
    \Gamma_{\rm C,max}^{\rm neutral} &=& 2~{\rm Im}(\omega_1), \nonumber \\
    &=&
    \sqrt{\frac{\pi}{32 \rm e}} \frac{m_{\rm e} c^2}{k_{\rm B}T_{\rm e}}
    \left( \frac{\omega_0}{\omega_{\rm c}} \right)^4
    \left( 1 + \frac{\omega_{\rm p}^2}{\omega_{\rm c}^2} \right)^{-1}
    \frac{a_{\rm e}^2 \omega_{\rm p}^2}{\omega_0}.~~~~
    \label{eq:gr_lin_n}
\end{eqnarray}
As in the case of the charged mode, we now consider a circularly polarized incident wave.  
From Eq.~(\ref{eq:ae_rel}), the maximum growth rate of the neutral mode 
for a circularly polarized incident wave, $\Gamma_{\rm C,max}^{\rm circ,neutral}$, is given by
\begin{eqnarray}
    \Gamma_{\rm C,max}^{\rm circ,neutral} =
    \sqrt{\frac{\pi}{8 \rm e}} \frac{m_{\rm e} c^2}{k_{\rm B}T_{\rm e}}
    \left( \frac{\omega_0}{\omega_{\rm c}} \right)^4
    \left( 1 + \frac{\omega_{\rm p}^2}{\omega_{\rm c}^2} \right)^{-1}
    \frac{\left(a_{\rm e}^{\rm circ}\right)^2 \omega_{\rm p}^2}{\omega_0} .\nonumber \\
    \label{eq:gr_circ_n}
\end{eqnarray}

According to Refs.~\cite{nishiura25a,nishiura25b}, the growth rates of the charged and neutral modes are maximized for backward scattering ($\nu = -1, \cos \theta_{kB} = \pm1$).
Furthermore, the wavenumbers corresponding to the maximum growth for the density fluctuation, $k_{\rm max}$, and for the scattered wave, $k_{1,{\rm max}}$, in both the charged and neutral modes are given by Eq.~(118) in Ref.~\cite{nishiura25b}:
\begin{eqnarray}
    k_{\rm max} &=& 2k_0 \left\{1 - \sqrt{ \frac{k_{\rm B} T_{\rm e}}{m_{\rm e} c^2} \left( 1 + \frac{\omega_{\rm p}^2}{\omega_{\rm c}^2} \right) } \right\}, \label{eq:kmax} \\
    k_{\rm 1,max} &=& k_{\rm max} - k_0, \nonumber \\
    &=& k_0 \left\{1 - 2\sqrt{ \frac{k_{\rm B} T_{\rm e}}{m_{\rm e} c^2} \left( 1 + \frac{\omega_{\rm p}^2}{\omega_{\rm c}^2} \right) } \right\}, \label{eq:k1max}
\end{eqnarray}
where $k_0$ is the wavenumber of the incident wave.

\section{Simulation Setup} \label{sec:setup}
\begin{figure}[htbp]
    \centering
    \includegraphics[width=1.0\linewidth]{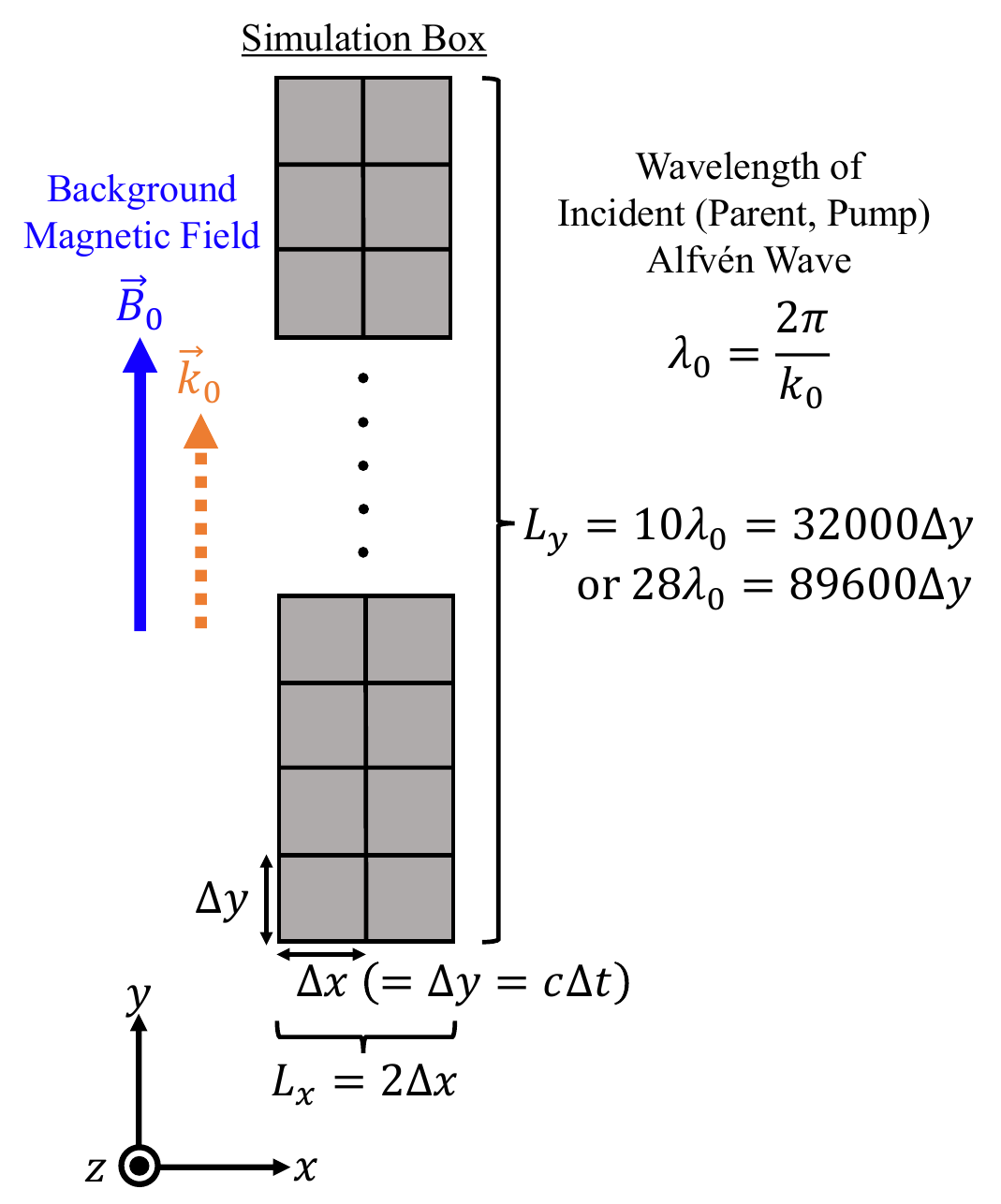}
    \caption{\RaggedRight Schematic of the simulation box. The fiducial box size is $L_x = 2\Delta x$ in the $x$ direction and $L_y = 10\lambda_0 = 32000\Delta y$ ($L_y = 28\lambda_0 = 89600\Delta y$) for the charged (neutral) mode simulations in the $y$ direction. The background magnetic field $\bm{B}_0$ and the incident Alfv\'en wave vector $\bm{k}_0$ are both aligned with the $y$ direction.}
    \label{fig:sim_setup}
\end{figure}
We perform PIC simulations using WumingPIC2D \cite{matsumoto24} to study the charged and neutral modes of induced Compton scattering in a magnetized electron-positron pair plasma with a circularly polarized incident Alfv\'en wave.  
The code employs a second-order shape function for computational macroparticles, the charge-conserving scheme of Ref.~\cite{esirkepov01}, and an implicit Maxwell solver without digital filtering \cite{ikeya15}.  
As a result, the simulations are free from the Courant-Friedrichs-Lewy constraint.  
In code units, the grid size in $x$ and $y$ directions, the speed of light, and the electron mass are unity ($\Delta x = \Delta y = 1, c=1, m_{\rm e}=1$).
We follow all components of both the electromagnetic fields $(E_x,E_y,E_z,B_x,B_y,B_z)$ and the particle velocities $(v_x/c,v_y/c,v_z/c)$.
The Boris particle pusher is adopted as the fiducial particle pusher \cite{birdsall91} (see Appendix~\ref{sec:nc}).

Figure~\ref{fig:sim_setup} shows the schematic of the simulation setup.  
The rectangular box lies in the {\it x--y} plane, with periodic boundaries applied in both $x$ and $y$ directions.
The box size in the $x$ direction is $L_x = N_x \Delta x= 2\Delta x$, where $2\Delta x$ is the minimum box size due to the code constraint. 
The box size in the $y$ direction is $L_y = N_y \Delta y =10\lambda_0 = 32000\Delta y$ for the fiducial charged mode simulations and $L_y = N_y \Delta y = 28\lambda_0 = 89600\Delta y$ for the fiducial neutral mode simulations, where $\lambda_0 = 2\pi/k_0$ is the incident wavelength.
$N_x$ and $N_y$ are the number of grids in $x$ and $y$ directions.
Since $L_y \gg L_x$, our simulations are effectively 1D simulations.
The grid spacing is $\Delta x = \Delta y = c\Delta t$. 
$\Delta t$ is the time step.  
To resolve the scattered wave number (Eq.~(\ref{eq:k1max})), $L_y$ should satisfy   
\begin{eqnarray}
    L_y > \frac{2\pi}{\left| \bm{k}_0 \right| - \left| \bm{k}_1 \right|} 
    \simeq \frac{\lambda_0}{2\sqrt{k_{\rm B}T_{\rm e}/(m_{\rm e}c^2)}},
\end{eqnarray}
where we assume $\omega_{\rm c} \gg \omega_{\rm p}$ (Eq.~(\ref{eq:wcwp})) \cite{ghosh22}.
Thus, $L_y/\lambda_0 \gtrsim 2.4~(\text{for}~\sqrt{k_{\rm B}T_{\rm e}/(m_{\rm e}c^2)}=0.21)$ and $L_y/\lambda_0 \gtrsim 13~(\text{for}~\sqrt{k_{\rm B}T_{\rm e}/(m_{\rm e}c^2)}=0.04)$ are required for the charged and neutral modes, respectively.
The charged (neutral) mode dominates for the case of $\sqrt{k_{\rm B}T_{\rm e}/(m_{\rm e}c^2)}=0.21~(0.04)$ under our simulation setup (see Table~\ref{tab:param}).
We fix $\omega_0/\omega_{\rm pe} = 0.9$.
The condition $\omega_0<\omega_{\rm pe}$ is satisfied
so that the incident wave is an Alfv\'en wave.
We adopt $|\omega_{\rm ce}|\Delta t = (\omega_{\rm pe}/\omega_0)\sqrt{\sigma_{\rm e}}\,\omega_0\Delta t \approx 0.1$.  
The electron magnetization parameter is $\sigma_{\rm e} = B_0^2/(4\pi n_0 m_{\rm e} c^2) = 2500$, with the background field $\bm{B}_0 = B_0 \hat{\bm{y}}$. 
We assume a uniform background density with equal electron and positron masses, and use $n_{\rm ppc}=100$ particles per cell for both electrons and positrons as the fiducial value. 
$n_{\rm ppc}$ is same as $n_0$ in simulations.

The initial electron and positron distributions are isotropic Maxwell-J\"uttner in the plasma rest frame, generated by the modified Swisdak reduction method \cite{zenitani24}.  
The thermal velocities are $\sqrt{k_{\rm B}T_{\rm e}/m_{\rm e}} = 0.21c$ and $0.04c$ for the charged and neutral mode-dominated calculations, respectively.  
In the simulation (lab) frame, electrons and positrons have bulk oscillatory motion (see Eq.~(\ref{eq:vs})), and the isotropic distribution is Lorentz-transformed accordingly with particle number corrections in each velocity bin applied \cite{zenitani15}.
\begin{table*}[htbp]
\renewcommand{\arraystretch}{1.3}
\newcolumntype{C}[1]{>{\centering\arraybackslash}p{#1}}
\newcommand{\mathcell}[1]{%
  \rule{0pt}{18pt}%
  \(\displaystyle #1\)%
  \raisebox{-12pt}{\rule{0pt}{4pt}}%
}
\centering
\caption{Simulation Parameters}
\begin{tabular}{C{1.2cm} C{1.3cm} C{2.5cm} C{1.1cm} C{1.8cm} C{2.0cm} C{2.4cm} C{1.6cm} C{2cm}}
\hline\hline
\multicolumn{9}{c}{\textbf{Common parameters:} 
\mathcell{\sigma_{\rm e} = \frac{B_0^2}{4\pi n_0 m_{\rm e} c^2} = 2500,~~ 
\frac{\omega_0}{\omega_{\rm pe}}=0.9,~~
\frac{\lambda_0}{\Delta x}=3200,~~
N_x=\frac{L_x}{\Delta x}=2}}  \\
\hline
Run & \mathcell{\sqrt{\frac{k_{\rm B} T_{\rm e}}{m_{\rm e} c^2}}}
    & \mathcell{\eta^{\rm circ}_{\rm inci}=\frac{B_{\rm inci}}{B_0}}
    & \mathcell{a^{\rm circ}_{\rm e}}
    & \mathcell{N_y=\frac{L_y}{\Delta y}}
    & \mathcell{n_{\rm ppc}~[\rm /cell]}
    & Particle Pusher
    & Mode
    & \mathcell{\frac{\Gamma_{\rm C,max}^{\rm circ,sim}}{\omega_0}} \\
\hline
Run 1  & \multirow{4}{*}{0.21}  & 0.1000 & 5.553 &\multirow{4}{*}{32000} & \multirow{4}{*}{100} & \multirow{4}{*}{Boris} & \multirow{4}{*}{Charged} & $9.392\times10^{-4}$ \\
Run 2  &                        & 0.1778 & 9.874 &                       &                      &       &                           & $2.743\times10^{-3}$ \\
Run 3  &                        & 0.3162 & 17.56 &                       &                      &       &                           & $9.339\times10^{-3}$ \\
Run 4  &                        & 0.5623 & 31.23 &                       &                      &       &                           & $4.013\times10^{-2}$ \\
\hline
Run 5  & 0.21                   & 0.3162 & 17.56 & 89600                 & 100                  & Boris & Charged                  & $9.308\times10^{-3}$ \\
\hline
Run 6  & \multirow{2}{*}{0.21} & \multirow{2}{*}{0.3162} &\multirow{2}{*}{17.56} &\multirow{2}{*}{32000} & 200 & \multirow{2}{*}{Boris} & \multirow{2}{*}{Charged} & $9.206\times10^{-3}$ \\
Run 7  &                       &                         &                       &                       & 400 &                        &                          & $8.636\times10^{-3}$ \\
\hline
Run 8  & \multirow{4}{*}{0.21} & 0.1000 & 5.553 &\multirow{4}{*}{32000} & \multirow{4}{*}{100} & \multirow{4}{*}{Vay} & \multirow{4}{*}{Charged} & $9.867\times10^{-4}$ \\
Run 9  &                       & 0.1778 & 9.874 &                       &                      &                      &                          & $3.226\times10^{-3}$ \\
Run 10 &                       & 0.3162 & 17.56 &                       &                      &                      &                          & $9.179\times10^{-3}$ \\
Run 11 &                       & 0.5623 & 31.23 &                       &                      &                      &                          & $4.030\times10^{-2}$ \\
\hline
Run 12 & \multirow{4}{*}{0.21} & 0.1000 & 5.553 &\multirow{4}{*}{32000} & \multirow{4}{*}{100} & \multirow{4}{*}{Higuera-Cary} & \multirow{4}{*}{Charged} & $9.249\times10^{-4}$ \\
Run 13 &                       & 0.1778 & 9.874 &                       &                      &                               &                          & $2.931\times10^{-3}$ \\
Run 14 &                       & 0.3162 & 17.56 &                       &                      &                               &                          & $9.058\times10^{-3}$ \\
Run 15 &                       & 0.5623 & 31.23 &                       &                      &                               &                          & $3.325\times10^{-2}$ \\
\hline
Run 16 & \multirow{4}{*}{0.04} & 0.1000 & 5.553 & \multirow{4}{*}{89600} & \multirow{4}{*}{100} & \multirow{4}{*}{Boris} & \multirow{4}{*}{Neutral} & $1.586\times10^{-3}$ \\
Run 17 &                       & 0.1778 & 9.874 &                        &                      &                        &                          & $6.010\times10^{-3}$ \\
Run 18 &                       & 0.3162 & 17.56 &                        &                      &                        &                          & $1.844\times10^{-2}$ \\
Run 19 &                       & 0.5623 & 31.23 &                        &                      &                        &                          & $5.123\times10^{-2}$ \\
\hline
Run 20 & 0.04                  & 0.3162 & 17.56 & 134400 & 100 & Boris & Neutral & $1.771\times10^{-2}$ \\
\hline\hline
\multicolumn{9}{c}{ Note: $\Delta x =\Delta y= c\Delta t,\ \omega_0 \Delta t < \omega_{\rm pe} \Delta t < \omega_{\rm c}\Delta t = \sqrt{\sigma_{\rm e}} \omega_{\rm pe}\Delta t \lesssim 0.1$}
\end{tabular}
\label{tab:param}
\end{table*}

The incident wave is a monochromatic, right-handed, circularly polarized Alfv\'en wave, initially given by  
\begin{eqnarray}
    \bm{B}_{\rm inci} 
    &=& \left( -B_{\rm inci} \sin(k_0 y),\ 0,\ B_{\rm inci} \cos(k_0 y) \right), \\
    \bm{E}_{\rm inci} 
    &=& \left( -\frac{\omega_0}{c k_0}B_{\rm inci} \cos(k_0 y),\ 0,\ 
    -\frac{\omega_0}{c k_0}B_{\rm inci} \sin(k_0 y) \right). \nonumber \\
    \label{eq:ep_bp}
\end{eqnarray}
The results are identical for right- and left-handed circular polarizations because our simulations consider an electron-positron pair plasma.
The dispersion relation of the incident Alfv\'en wave is \cite{matsukiyo03}  
\begin{eqnarray}
    \left( \frac{ck_0}{\omega_0} \right)^2 
    = 1 - \sum_{\rm s} 
    \frac{\omega_{\rm ps}^2}{\omega_0 \left( \gamma_{\rm s}\omega_0 + \omega_{\rm cs} \right)}, 
    \label{eq:disp}
\end{eqnarray}
where the index ${\rm s}$ denotes particle species (electrons and positrons).  
Here, $\omega_{\rm ps} = \sqrt{4\pi n_0 e^2/m_{\rm e}} = \omega_{\rm pe} = \omega_{\rm p}/\sqrt{2}$ and $\omega_{\rm cs} = \pm eB_0/(m_{\rm e}c) = \pm \omega_{\rm c}$ are plasma and cyclotron frequencies, respectively, for each species.
The plus (minus) sign corresponds to positrons (electrons).  
In the simulation frame, in order to self-consistently determine the current associated with the incident Alfv\'en wave, the initial transverse velocity of each species is given by \cite{matsukiyo03}
\begin{eqnarray}
    \frac{\bm{v}_{\rm s}}{c} &=& 
    - \frac{\omega_0}{ck_0} 
    \frac{\eta^{\rm circ}_{\rm inci} \,\omega_{\rm cs}}
    {\gamma_{\rm s}\omega_0 + \omega_{\rm cs}} 
    \frac{\bm{B}_{\rm inci}}{B_{\rm inci}}, 
    \label{eq:vs} \\
    \gamma_{\rm s} &=& \frac{1}{\sqrt{1 - \left( v_{\rm s}/c \right)^2}}, \\
    \eta^{\rm circ}_{\rm inci} &=&  \frac{B_{\rm inci}}{B_0}. 
    \label{eq:eta}
\end{eqnarray}
$\eta^{\rm circ}_{\rm inci}$ is the relative amplitude of the incident Alfv\'en wave and $v_{\rm s}/c \approx \eta^{\rm circ}_{\rm inci}\omega_0/(ck_0)$ under the condition that $\omega_0 < \omega_{\rm p} \ll \omega_{\rm c}$ and the particle velocity is nonrelativistic.
The three unknowns, $\omega_0\Delta t$, $v_{+}/c$, and $v_{-}/c$, are determined by solving Eq.~(\ref{eq:disp}) together with Eq.~(\ref{eq:vs}) for both electrons and positrons.  
The zeroth-order longitudinal velocity is set to zero since a circularly polarized Alfv\'en wave does not induce a pressure gradient along the background field ($y$ direction in our simulations).  
Using Eqs.~(\ref{eq:ae_circ}), (\ref{eq:ep_bp}), and (\ref{eq:eta}), the strength parameter for the circularly polarized wave $a_{\rm e}^{\rm circ}$ is expressed as  
\begin{eqnarray}
    a_{\rm e}^{\rm circ} 
    = \frac{eE_{\rm inci}}{m_{\rm e}c\omega_0}
    = \frac{eB_{\rm inci}}{m_{\rm e}c^2k_0}  
    = \eta^{\rm circ}_{\rm inci} \frac{\omega_{\rm c}}{\omega_0} \frac{v_{\rm A}}{c}, \label{eq:ae_circ2}\nonumber \\
\end{eqnarray}
where the relativistic Alfv\'en velocity is  
\begin{eqnarray}
    v_{\rm A} = c\left(1 + \frac{\omega_{\rm p}^2}{\omega_{\rm c}^2}\right)^{-1/2}.
\end{eqnarray}
The numerical parameters are summarized in Table~\ref{tab:param}, and the numerical convergence is presented in Appendix~\ref{sec:nc}.

\section{Simulation Results} \label{sec:result}
\subsection{Linear growth rate of charged and neutral modes}
\begin{figure*}[htbp]
    \centering
    \includegraphics[width=1.0\linewidth]{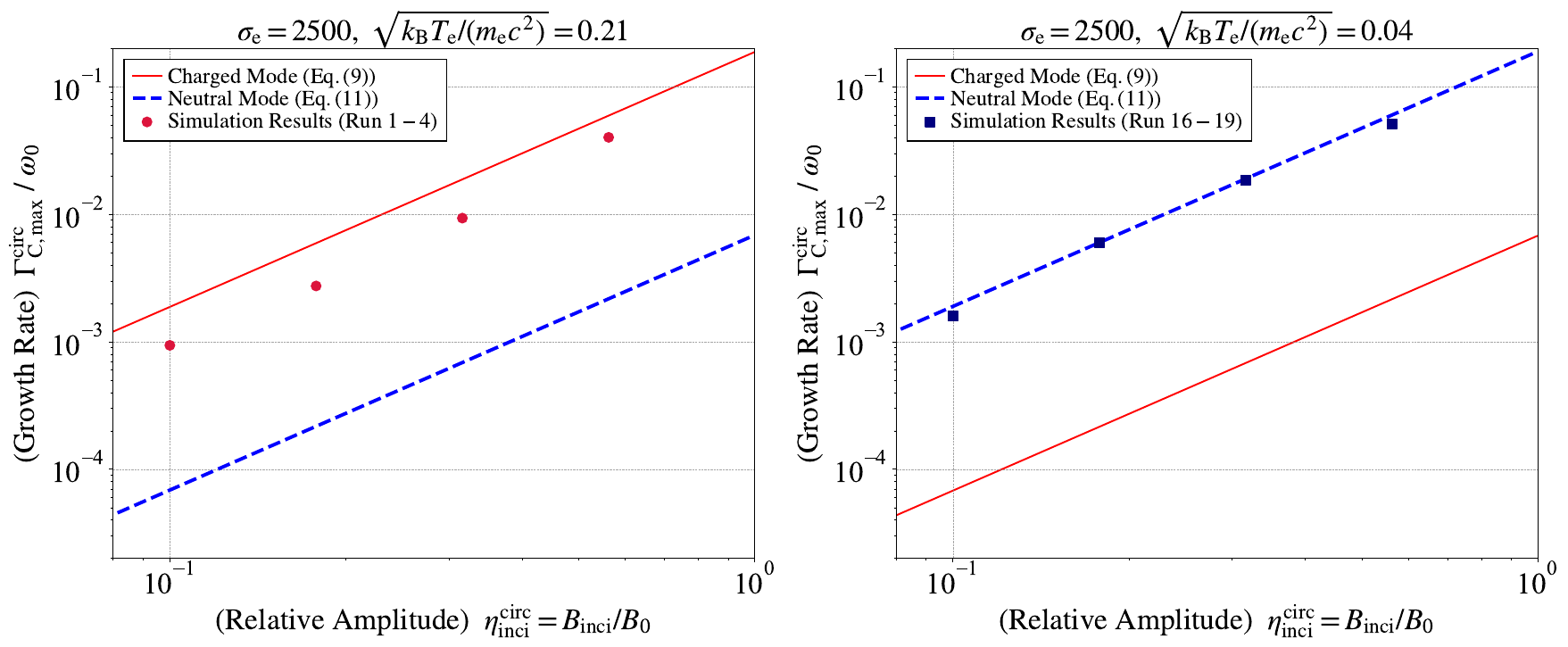}
    \caption{\RaggedRight 
    Linear growth rate $\Gamma_{\rm C,max}^{\rm circ}$ as a function of the relative amplitude of the circularly polarized incident Alfv\'en wave $\eta^{\rm circ}_{\rm inci}$.  
    The vertical axis shows the linear growth rate $\Gamma_{\rm C,max}^{\rm circ}$ normalized by the incident wave frequency $\omega_0$.  
    The red solid and blue dashed lines indicate the analytical growth rates of the charged mode $\Gamma_{\rm C,max}^{\rm circ,charged}$ (Eq.~(\ref{eq:gr_circ_c})) and the neutral mode, $\Gamma_{\rm C,max}^{\rm circ,neutral}$ (Eq.~(\ref{eq:gr_circ_n})), respectively.  
    Red circles (Runs~1-4) and blue squares (Runs~16-19) denote simulation results.  
    The left panel corresponds to $\sqrt{k_{\rm B}T_{\rm e}/(m_{\rm e}c^2)}=0.21$, where the charged mode dominates, and the right panel corresponds to $\sqrt{k_{\rm B}T_{\rm e}/(m_{\rm e}c^2)}=0.04$, where the neutral mode dominates.
    }
    \label{fig:gmax}
\end{figure*}
Figure~\ref{fig:gmax} shows the linear growth rate of the circularly polarized wave 
$\Gamma_{\rm C,max}^{\rm circ}$ as a function of the amplitude of the incident 
circularly polarized Alfv\'en wave $\eta^{\rm circ}_{\rm inci}$ (Eq.~(\ref{eq:eta})).
The vertical axis is normalized by the incident wave frequency $\omega_0$.  
The red solid line denotes the analytical linear growth rate of the charged mode, $\Gamma_{\rm C,max}^{\rm circ,charged}$ (Eq.~(\ref{eq:gr_circ_c})), while the blue dashed line represents that of the neutral mode, $\Gamma_{\rm C,max}^{\rm circ,neutral}$ (Eq.~(\ref{eq:gr_circ_n})).  
Simulation results are indicated by red circles (Runs~1--4) and blue squares (Runs~16--19), and are also summarized in Table~\ref{tab:param}.  
The left panel corresponds to $\sqrt{k_{\rm B}T_{\rm e}/(m_{\rm e}c^2)}=0.21$, where the charged mode dominates, and the right panel corresponds to $\sqrt{k_{\rm B}T_{\rm e}/(m_{\rm e}c^2)}=0.04$, where the neutral mode dominates.  
In both cases, the simulations reproduce the predicted $(\eta^{\rm circ}_{\rm inci})^2$ scaling.  
For the charged mode (left panel), the growth rate obtained from the simulations is smaller by about a factor of two compared with the analytical estimate. 
For the neutral mode (right panel), the simulation results are in good agreement with the analytical estimate (Eq.~(\ref{eq:gr_circ_n})).

In the charged mode case, the slight discrepancy between the simulation results and the analytical estimate can be attributed to two main effects:
\begin{enumerate}
\item \textbf{Thermal velocity}: In the charged mode runs, a relatively large thermal 
velocity, $\sqrt{k_{\rm B}T_{\rm e}/(m_{\rm e}c^2)}=0.21$, was used due to computational time constraints.  
This increases the frequency difference between the scattered and incident waves 
($\omega_1$ and $\omega_0$), reducing the accuracy of the assumption $\omega_1 \sim \omega_0$ in the analytical estimate (see Eq.~(74) in \cite{nishiura25a}).
As a result, both the ponderomotive potential and the dispersion relation of the scattered wave 
(Eqs.~(58) and (95) in Ref.~\cite{nishiura25a}) are modified, which affects the linear growth rate.

\item \textbf{Maximization procedure}: Ref.~\cite{nishiura25a} expressed the growth rate 
as a function of $\zeta = \omega/(k_\parallel v_{\rm th})$ and maximized it only with respect to $\zeta$ 
(see Eq.~(67) therein).  
However, since $\omega_1$ and $k$ themselves also depend on $\zeta$, the growth rate should be 
maximized taking these dependencies into account.
This effect can be negligible for the low thermal velocity.
This is because the $\zeta$ dependency of $\omega_1$ and $k$ becomes weak and $\omega_1\approx \omega_0$ and $k\approx 2k_0$ for the low thermal velocity case.
Including this effect along with the thermal velocity contribution (effect 1 above), the value giving the maximum 
growth rate is deviated from $\zeta = -1/\sqrt{2}$ that is used in Ref.~\cite{nishiura25a}.
\end{enumerate}

\subsection{Time evolution of the power of incident and scattered waves}
\begin{figure*}[htbp]
    \centering
    \includegraphics[width=1.0\linewidth]{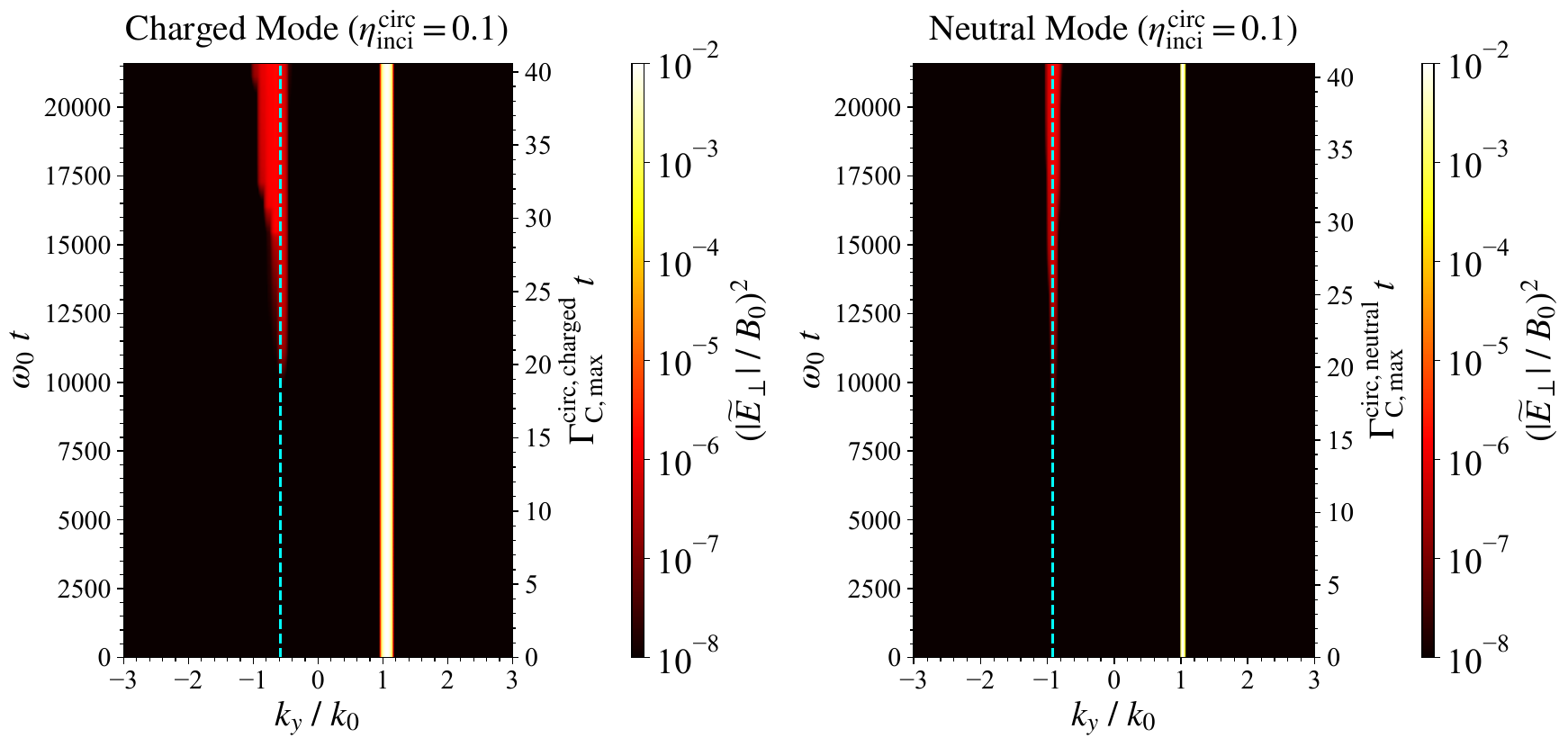}
    \caption{\RaggedRight 
    Time evolution of the transverse wave power normalized by the background magnetic field strength $(|\widetilde{E}_\perp(k_y,t)|/B_0)^2$.
    The left panel shows results for the charged mode (Run~1: $\eta^{\rm circ}_{\rm inci}=0.1$) and the right panel shows results for the neutral mode (Run~16: $\eta^{\rm circ}_{\rm inci}=0.1$).  
    The horizontal axis shows the wavenumber in the $y$ direction $k_y$ normalized by the wavenumber of the incident wave $k_0$. 
    The left vertical axis in both panels indicates the time normalized by the incident wave frequency $\omega_0$, while the right vertical axis in the left and right panel is normalized by the maximum growth rate of the charged and neutral modes, ($\Gamma_{\rm C,max}^{\rm circ,charged}$ (Eq.~(\ref{eq:gr_circ_c})) and $\Gamma_{\rm C,max}^{\rm circ,neutral}$ (Eq.~(\ref{eq:gr_circ_n}))), respectively.
    The color shows $(|\widetilde{E}_\perp(k_y,t)|/B_0)^2$.
    The vertical cyan line denotes the analytical estimate of the wavenumber of the backward-scattered wave, $k_y/k_0 = -k_{\rm 1,max}/k_0$ (see Eq.~(\ref{eq:k1max})).
    }
    \label{fig:ky_wt}
\end{figure*}
Next, we consider the time evolution of the power of the transverse waves (incident and scattered waves).  
For each snapshot of the field data, we Fourier transform the transverse electric field,
\begin{eqnarray}
    E_\perp(y,t) = \langle E_z \rangle_x(y,t) - \mathrm{i}\langle E_x \rangle_x(y,t), \label{eq:eperp}
\end{eqnarray}
in the $y$ direction to decompose the forward- and backward-propagating waves \cite{terasawa86,amano13}, which correspond to positive and negative wavenumbers, respectively.
$\langle E_x \rangle_x$ and $\langle E_z \rangle_x$ are the $x$ and $z$ components of the transverse wave averaged over the $x$ direction, respectively.
The discrete Fourier transform is calculated by using the \texttt{numpy.fft.fft} function of the NumPy library~\citep{numpy}.
The discrete Fourier transformation of $E_\perp(y,t)$ is given as follows: 
\begin{eqnarray}
    \widetilde{E}_\perp(k_{y,l},t) = \frac{1}{N_y} \sum_{m=0}^{N_y -1} E_\perp(y_m,t)\exp \left( - 2\pi{\mathrm i} \frac{ml}{N_y} \right)&&\nonumber\\
    (l=0,...,N_y-1),&& \nonumber\\
\end{eqnarray}
where $k_{y,l} = l\Delta k_y = 2\pi l/L_y = 2\pi l/(N_y \Delta y),~y_m = m\Delta y$.
To visualize the spectrum with negative and positive wavenumbers centered at zero, the FFT output was shifted using \texttt{numpy.fft.fftshift}.

Figure~\ref{fig:ky_wt} shows the time evolution of the Fourier power of the transverse waves $(|\widetilde{E}_\perp|/B_0)^2$.
The left panel corresponds to the charged mode with $\eta^{\rm circ}_{\rm inci}=0.1$ (Run~1), 
and the right panel corresponds to the neutral mode with $\eta^{\rm circ}_{\rm inci}=0.1$ (Run~16).  
The horizontal axis is the wavenumber in the $y$ direction $k_y$ normalized by the wavenumber of the incident wave $k_0$.  
The left vertical axis in both panels represents time normalized by the incident wave frequency 
$\omega_0$ while the right vertical axis represents time normalized by the maximum growth rate 
of the corresponding mode, i.e., $\Gamma_{\rm C,max}^{\rm circ,charged}$ 
(Eq.~(\ref{eq:gr_circ_c})) for the left panel and $\Gamma_{\rm C,max}^{\rm circ,neutral}$ 
(Eq.~(\ref{eq:gr_circ_n})) for the right panel.  
The color scale indicates the Fourier power of the transverse waves, $(|\widetilde{E}_\perp|/B_0)^2$.
The vertical cyan line shows the analytically estimated wavenumber of the backward-scattered 
wave corresponding to the maximum growth (Eq.~(\ref{eq:k1max})). 
At $\omega_0 t = 0$, only the incident wave ($k_y/k_0 = 1$) is present.  
As time progresses, the power in the negative $k_y$ region grows due to induced Compton scattering.
Furthermore, in both panels, the wavenumber of the scattered wave is in good agreement with the analytical estimate, $-k_{\rm 1,max}/k_0$ (Eq.~(\ref{eq:k1max})).
The resolution of $k_y$ for fiducial charged mode simulations (Run 1--4) and fiducial neutral mode simulations (Run 16--19) is given by $\Delta k_y/k_0 =  2\pi/(k_0 L_y) = \lambda_0/L_y = 0.1$ and $\Delta k_y/k_0 \approx 0.036$, respectively.
For the case with $\eta_{\rm inci}^{\rm circ}=0.3162$, we performed both the charged mode (Run~5) and neutral mode (Run~20) simulations using a larger $L_y$ (a smaller $\Delta k_y$) than $L_y$ used in the fiducial setup.
For the case of $\eta_{\rm inci}^{\rm circ}=0.3162$, when comparing the two $L_y$ cases, 
although the deviation of the scattered-wave power between the fiducial and smaller $\Delta k_y/k_0$ cases is within a factor of two for both the charged mode (Runs~3 and~5) and the neutral mode (Runs~18 and~20), the growth rates for both the charged mode (Runs 3 and 5) and neutral mode (Runs 18 and 20) are nearly identical.

\begin{figure*}[htbp]
    \centering
    \includegraphics[width=1.0\linewidth]{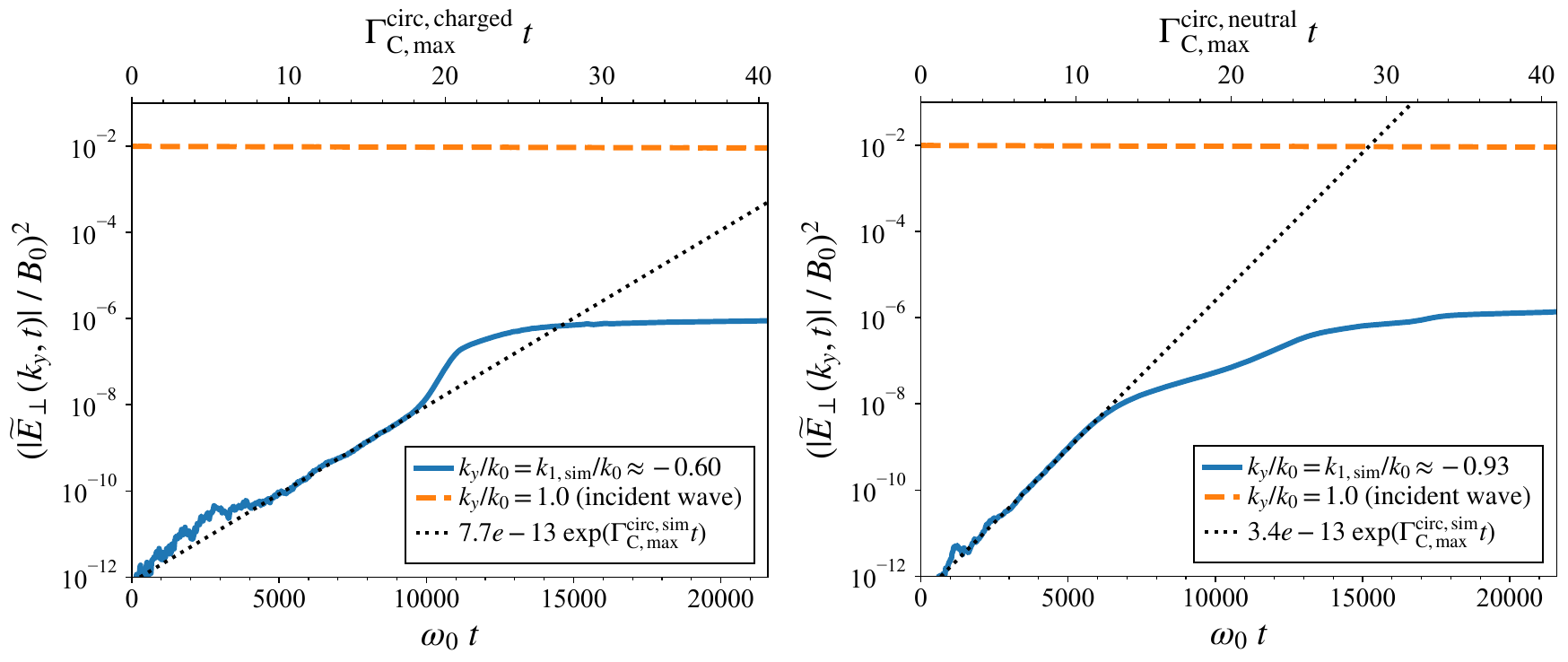}
    \caption{\RaggedRight 
    Time evolution of the Fourier power of the incident wave (orange dashed line) and the scattered wave (blue solid line) for the maximum growth (fastest-growing scattered wave).
    The left panel shows results for the charged mode (Run~1: $\eta^{\rm circ}_{\rm inci}=0.1$) and the right panel shows results for the neutral mode (Run~16: $\eta^{\rm circ}_{\rm inci}=0.1$).  
    The bottom horizontal axis in both panels is the time normalized by the incident wave frequency $\omega_0$.
    The top horizontal axis in the left and right panel is normalized by the analytical growth rate of the charged mode $\Gamma_{\rm C,max}^{\rm circ,charged}$ (Eq.~(\ref{eq:gr_circ_c})) and the neutral mode $\Gamma_{\rm C,max}^{\rm circ,neutral}$ (Eq.~(\ref{eq:gr_circ_n})), respectively. 
    The vertical axis shows the Fourier power of the transverse wave, normalized by the square of the background magnetic field strength $(|\widetilde{E}_\perp(k_y,t)|/B_0)^2$.
    The black dotted line shows the fitted curve, using the function $a \exp(\Gamma_{\rm C,max}^{\rm circ,sim} t)$.
    }
    \label{fig:wt_power}
\end{figure*}
Next, we extract the backward-propagating wave with the largest growth rate from 
Fig.~\ref{fig:ky_wt}.  
We focus on the region $k_y < 0$, which corresponds to the backward-propagating wave.  
Figure~\ref{fig:ky_wt} shows the time evolution of the scattered-wave power with a different wavenumber.
For each scattered wave, the linear growth phase is fitted by the function $a \exp(\Gamma t)$.
The scattered wave yielding the largest growth rate $\Gamma$ from this fitting is identified as the \textit{fastest-growing scattered wave}, 
and its growth rate is taken as the maximum growth rate obtained from the simulation.

Figure~\ref{fig:wt_power} presents the time evolution of the power of the incident wave and the fastest-growing scattered wave. 
The bottom horizontal axis in both panels represents time normalized by the incident wave frequency $\omega_0$, 
while the top horizontal axis represents time normalized by the analytical linear growth rate of the corresponding mode, 
 $\Gamma_{\rm C,max}^{\rm circ,charged}$ (Eq.~(\ref{eq:gr_circ_c})) for the left panel and 
$\Gamma_{\rm C,max}^{\rm circ,neutral}$ (Eq.~(\ref{eq:gr_circ_n})) for the right panel. 
The vertical axis shows the Fourier power of the transverse wave, $(|\widetilde{E}_\perp(k_y,t)|/B_0)^2$.
The left panel corresponds to the charged mode with $\eta^{\rm circ}_{\rm inci} = 0.1$ (Run~1), 
and the right panel corresponds to the neutral mode with $\eta^{\rm circ}_{\rm inci} = 0.1$ (Run~16).  
The blue solid line indicates the Fourier power of the fastest-growing scattered wave.
The orange dashed line shows the Fourier power of the incident wave.
For the charged (neutral) mode, the wavenumber of the fastest-growing scattered wave, $k_y/k_0=k_{\rm 1,sim}/k_0$, is approximately $-0.60$ ($-0.93$).  
The black dotted line shows the fitted curve, using the function $a \exp(\Gamma_{\rm C,max}^{\rm circ,sim} t)$.  
The maximum growth rates from all simulations, $\Gamma_{\rm C,max}^{\rm circ,sim} / \omega_0$, are summarized in Table~\ref{tab:param}.
As shown in Fig.~\ref{fig:wt_power}, after the linear growth phase terminates, the scattering appears to saturate in the nonlinear stage. 
The interpretation of this nonlinear behavior is discussed in Sec.~\ref{ssec:saturation}.

\subsection{Time evolution of the density fluctuation}
\begin{figure*}[htbp]
    \centering
    \includegraphics[width=1.0\linewidth]{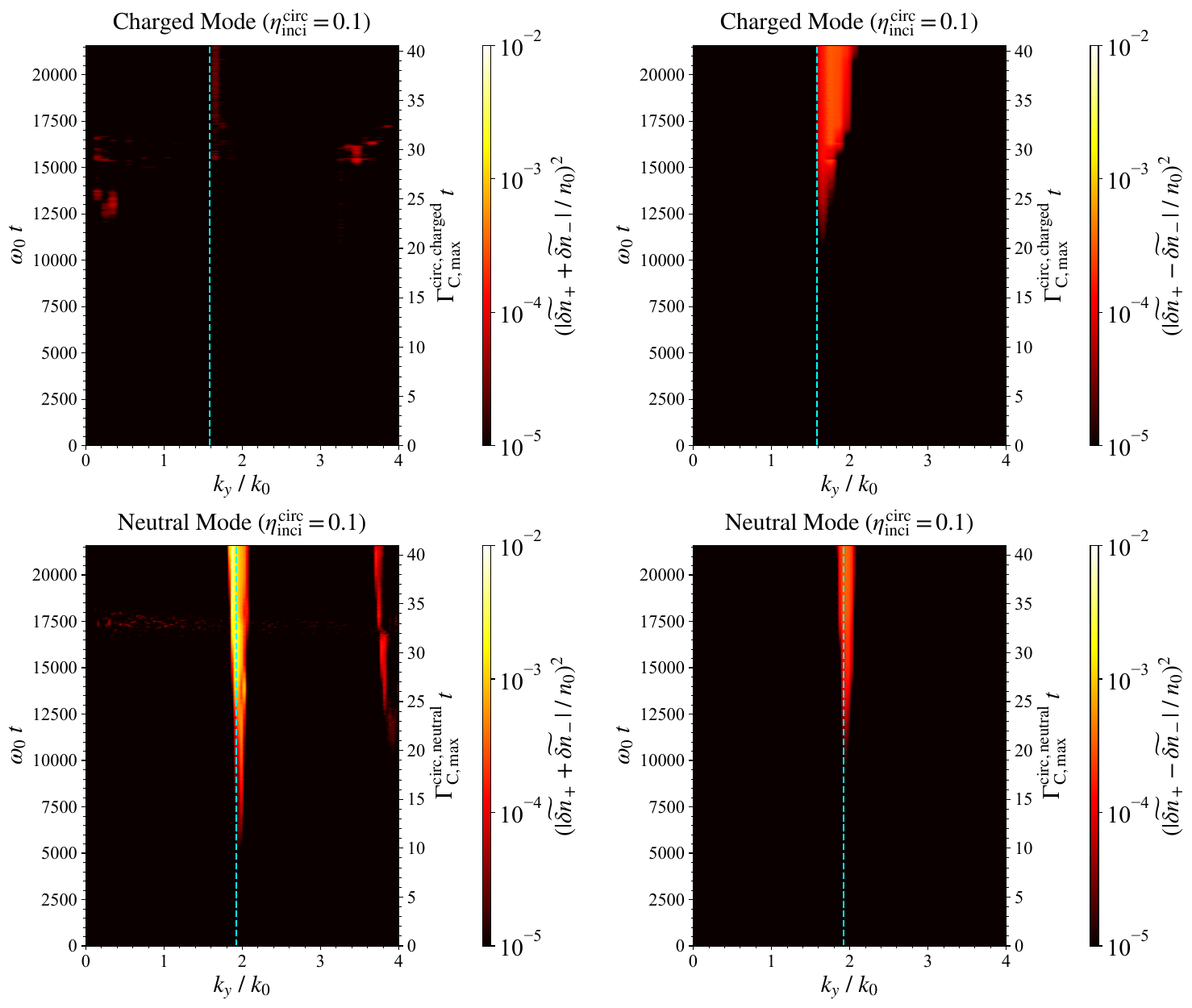}
    \caption{\RaggedRight 
    Time evolution of the Fourier power of the sum and difference of the positron and electron density fluctuations, $(|\widetilde{\delta n}_{\rm +} + \widetilde{\delta n}_{\rm -}|/n_0)^2$ and $(|\widetilde{\delta n}_{\rm +} - \widetilde{\delta n}_{\rm -}|/n_0)^2$, respectively (color-coded).
    The top two panels show results for the charged mode (Run~1: $\eta^{\rm circ}_{\rm inci} = 0.1$), while the bottom two panels show results for the neutral mode (Run~16: $\eta^{\rm circ}_{\rm inci} = 0.1$).  
    The left and right panels correspond to $(|\widetilde{\delta n}_{\rm +} + \widetilde{\delta n}_{\rm -}|/n_0)^2$ and $(|\widetilde{\delta n}_{\rm +} - \widetilde{\delta n}_{\rm -}|/n_0)^2$, respectively.
    The horizontal axis shows the wavenumber in the $y$ direction $k_y$ normalized by the wavenumber of the incident wave $k_0$.  
    The left vertical axis in all panels is the time normalized by the incident wave frequency $\omega_0$.  
    The right vertical axis in the top (bottom) panels is the time normalized by the maximum growth rate of the charged (neutral) mode $\Gamma_{\rm C,max}^{\rm circ,charged}$ (Eq.~(\ref{eq:gr_circ_c})) ($\Gamma_{\rm C,max}^{\rm circ,neutral}$ (Eq.~(\ref{eq:gr_circ_n}))).  
    The vertical cyan line shows the analytically estimated wavenumber of the density fluctuation (Eq.~(\ref{eq:kmax})).
    }
    \label{fig:kt_wt_dn}
\end{figure*}
We identify whether the growing mode corresponds to the charged or neutral mode by examining the density fluctuations of the electron-positron pair plasma.  
In the charged mode, the ponderomotive force drives electrons and positrons to accumulate at different spatial locations (see the left panel of Fig.~\ref{fig:mode}).  
Consequently, the sum of the density fluctuations, $\delta n_+ + \delta n_-$, vanishes, 
while the difference, $\delta n_+ - \delta n_-$, remains finite.  
Here, $\delta n_\pm = n_\pm - n_0$, where $n_\pm$ denotes the positron or electron density. 
In contrast, in the neutral mode, the ponderomotive force drives electrons and positrons to accumulate at the same spatial locations (see the right panel of Fig.~\ref{fig:mode}). 
In this case, the sum of the density fluctuations, $\delta n_+ + \delta n_-$, remains finite, while the difference, $\delta n_+ - \delta n_-$, vanishes. 
Therefore, by evaluating the Fourier power of the sum and difference of the density fluctuation, one can identify whether the excited mode corresponds to the charged or neutral mode.
We Fourier transform $(\delta n_+ + \delta n_-)/n_0$ and $(\delta n_+ - \delta n_-)/ n_0$ in the $y$ direction.
The Fourier transformation of $\delta n_\pm$ is calculated as follows:
\begin{eqnarray}
    \widetilde{\delta n}_\pm(k_{y,l},t) = \frac{2}{N_y} \sum_{m=0}^{N_y -1} \delta n_\pm(y_m,t)\exp \left( - 2\pi{\mathrm i} \frac{ml}{N_y} \right)&&\nonumber\\
    (l=0,...,N_y-1),&& \nonumber\\
    \label{eq:dn}
\end{eqnarray}
Similar to $\widetilde{E}_\perp$, since the number of samples of $\delta n_\pm$ is $N_y$, the normalization of Fourier transformation is $1/N_y$.
Contrary to $\widetilde{E}_\perp$, since $\delta n_\pm$ is real number, the Fourier-transformed component $\widetilde{\delta n}_\pm$ is symmetric with respect to $k_y=0$.
Here, we consider the one-side spectrum of $\widetilde{\delta n}_\pm$ ($\widetilde{\delta n}_\pm$ in the positive $k_y$ region).
To make the connection between the amplitude of $\widetilde{\delta n}_\pm$ and $\delta n_\pm$, the spectral components are multiplied by a factor of two.

Figure~\ref{fig:kt_wt_dn} shows the time evolution of the Fourier power of the sum and difference of the positron and electron density fluctuations, $ (|\widetilde{\delta n}_+ + \widetilde{\delta n}_- | / n_0)^2$ and $(|\widetilde{\delta n}_+ - \widetilde{\delta n}_-| / n_0)^2$, respectively.  
The top two panels correspond to the charged mode with $\eta^{\rm circ}_{\rm inci} = 0.1$ (Run~1), and the bottom two panels correspond to the neutral mode with $\eta^{\rm circ}_{\rm inci} = 0.1$ (Run~16). 
The left and right two panels show the time evolution of $(|\widetilde{\delta n}_+ + \widetilde{\delta n}_- | / n_0)^2$ and $(|\widetilde{\delta n}_+ - \widetilde{\delta n}_- | / n_0)^2$, respectively.
The horizontal axis is the wave number in the $y$ direction $k_y$ normalized by the wavenumber of the incident wave $k_0$.  
The left vertical axis represents time normalized by the incident wave frequency $\omega_0$ while the right vertical axis represents time normalized by the maximum growth rate of the corresponding mode, i.e., $\Gamma_{\rm C,max}^{\rm circ,charged}$ (Eq.~(\ref{eq:gr_circ_c})) for the left panel and $\Gamma_{\rm C,max}^{\rm circ,neutral}$ (Eq.~(\ref{eq:gr_circ_n})) for the right panel. 
The color indicates the Fourier power of the corresponding quantity ($(|\widetilde{\delta n}_+ + \widetilde{\delta n}_- | / n_0)^2$ or $ (|\widetilde{\delta n}_+ - \widetilde{\delta n}_-| / n_0)^2$).
In the top two panels, corresponding to the charged mode (Run~1), the power of $(|\widetilde{\delta n}_+ + \widetilde{\delta n}_- | / n_0)^2$ is almost absent, while that of $(|\widetilde{\delta n}_+ - \widetilde{\delta n}_- | / n_0)^2$ remains strong, in agreement with theoretical expectations.  
This feature is observed in all charged mode simulations conducted in this study (Runs~1--15), confirming that the charged mode is indeed excited as expected.  
On the other hand, in the bottom two panels for the neutral mode (Run~16), the power of $(|\widetilde{\delta n}_+ + \widetilde{\delta n}_- | / n_0)^2$ remains strong, while that of $(|\widetilde{\delta n}_+ - \widetilde{\delta n}_- | / n_0)^2$ is weak, also consistent with theoretical expectations.
Here, $(|\widetilde{\delta n}_+ - \widetilde{\delta n}_-| / n_0)^2$ is approximately one order of magnitude smaller than $(|\widetilde{\delta n}_+ + \widetilde{\delta n}_-| / n_0)^2$.
The expected reason why $(|\widetilde{\delta n}_+ - \widetilde{\delta n}_-| / n_0)^2$ remains finite in the neutral mode is because the ponderomotive force prevent the Debye screening.
In plasma, the charge separation with the frequency below the plasma frequency is suppressed by the Debye screening.
Hence, the density fluctuation that relates $(|\widetilde{\delta n}_+ - \widetilde{\delta n}_-| / n_0)^2$ is suppressed by the Debye screening.
However, since the ponderomotive force in magnetized plasma has a (minor) charge-dependent term (see third term of Eq.~(14) in Ref.~\cite{nishiura25a}), electrons and positrons attempt to move towards the opposite direction each other, which leads to the charge separation. 
Therefore, $(|\widetilde{\delta n}_+ - \widetilde{\delta n}_-| / n_0)^2$ remains finite because the ponderomotive force that generates the charge separation counteracts the Debye screening that suppresses the charge separation.
The behavior of strong $(|\widetilde{\delta n}_+ + \widetilde{\delta n}_-| / n_0)^2$ and weak $(|\widetilde{\delta n}_+ - \widetilde{\delta n}_-| / n_0)^2$ is observed in all neutral mode simulations (Runs~16--18), confirming that the neutral mode is realized as anticipated.
The vertical cyan line shows the analytically estimated wavenumber of the density fluctuation (Eq.~(\ref{eq:kmax})), which is consistent with the simulation results.
Therefore, these results demonstrate that both the charged and neutral modes expected for induced Compton scattering in a magnetized electron-positron pair plasma are successfully reproduced.

\subsection{Saturation mechanism of induced Compton scattering}\label{ssec:saturation}
In this paper, we have mainly focused on the linear growth stage. Nevertheless, in what follows we discuss a plausible mechanism of the nonlinear saturation. All simulation runs listed in Table~\ref{tab:param} exhibit a common saturation behavior. As shown in Fig.~\ref{fig:wt_power}, the scattered wave stops growing after the linear growth stage of induced Compton scattering, whereas the energy of the incident wave remains almost unchanged. We refer to this behavior as \emph{partial scattering}, because only a fraction of the incident-wave energy is transferred to the scattered wave. In contrast, one may also expect \emph{full scattering} in which most of the incident-wave energy is converted into the scattered wave. A detailed investigation of this regime will be presented in a companion paper (Nishiura {\it et al.} in prep.).

We consider that induced Compton scattering saturates once the scattered-wave energy density grows up to
\begin{equation}
    \varepsilon_{\mathrm{scat,max}}\sim \frac{1}{2}n_{0}m_{\mathrm{e}}v_{\mathrm{th}}v_{\mathrm{A}}.
    \label{eq:scattered_wave_energy_growth_saturate}
\end{equation}
We show that this condition corresponds to the case in which the energy transferred to the plasma through scattering becomes comparable to the internal energy density,
\begin{equation}
    \varepsilon_{\mathrm{th}}\equiv 2n_{0}\cdot\frac{1}{2}m_{\mathrm{e}} v_{\mathrm{th}}^{2},
    \label{eq:thermal_energy_density}
\end{equation}
where the one-dimensional thermal velocity is
\begin{equation}
    v_{\mathrm{th}}\equiv \sqrt{\frac{k_{\mathrm{B}}T_{\mathrm{e}}}{m_{\mathrm{e}}}}.
\end{equation}

The saturation condition in Eq.~\eqref{eq:scattered_wave_energy_growth_saturate} can be motivated by energy conservation in a single scattering event. Although the phenomenon under consideration is classical, it is helpful to describe a single scattering step in quantum-mechanical terms, namely by introducing $\hbar$ and considering the exchange of energy and momentum among the quanta of the incident wave, scattered wave, and density wave. We estimate the ratio between the energy gained by the scattered wave and the energy transferred to the plasma per step. The change in the photon energy is
\begin{equation}
        \delta\epsilon_{\gamma}=\hbar(\omega_1-\omega_0)
        \sim-2\hbar\omega_0\frac{v_{\mathrm{th}}}{v_{\mathrm{A}}},
    \label{eq:1photon_variation}
\end{equation}
where we used Eq.~\eqref{eq:k1max}, and $\omega_0\simeq k_0 v_{\mathrm{A}}$ and $\omega_1\simeq k_1 v_{\mathrm{A}}$ since both the incident and scattered waves are Alfv\'en waves. Energy conservation then gives the energy transferred to the density wave as $\delta\epsilon_{\mathrm{e}}=-\delta\epsilon_{\gamma}=\hbar\omega=\hbar(\omega_0-\omega_1)$. The ratio of the scattered photon energy to the particle energy gain per step is then obtained from Eq.~\eqref{eq:1photon_variation} and Eq.~\eqref{eq:k1max} as
\begin{equation}
   \frac{\hbar\omega_1}{\hbar(\omega_0-\omega_1)}=\frac{1}{2}\frac{v_{\mathrm{A}}}{v_{\mathrm{th}}}+\mathcal{O}(1).
    \label{eq:ratio_energy_saturation}
\end{equation}

Nonlinear saturation occurs when the distribution function is sufficiently modified near the phase velocity of the beat wave, $v_{\mathrm{ph}}\sim v_{\mathrm{th}}$, and a plateau is formed \citep{sagdeev69,melrose17}. We confirm this behavior in part in our simulations, and a more detailed analysis will be presented in a forthcoming paper. This saturation is expected because the three-wave resonance driven by the ponderomotive potential $\phi_{\mathrm{p}}^{\pm}$ involves particles with velocities $v\sim v_{\mathrm{th}}$. In the Vlasov equation (see Eq.~(10) in Ref.~\citep{nishiura25a}), the ponderomotive force enters through the term $-(1/m_{\mathrm{e}})\,\bm{\nabla}\phi_{\mathrm{p}}^{\pm}\cdot(\partial f_{\pm}/\partial\bm{v})$. Once a plateau develops in the resonant region, the derivative $\partial f_{\pm}/\partial\bm{v}$ becomes small. The ponderomotive force then becomes inefficient at driving the instability, and the scattered wave stops growing. As an alternative viewpoint, it is well known that Landau resonance is suppressed once a plateau forms.

Finally, we estimate the maximum energy density of the scattered wave that can be reached before saturation. Flattening the resonant part of the distribution requires an energy of order $\varepsilon_{\mathrm{th}}$. Multiplying $\varepsilon_{\mathrm{th}}$ by the ratio in Eq.~\eqref{eq:ratio_energy_saturation} and then substituting Eq.~\eqref{eq:thermal_energy_density} yields Eq.~\eqref{eq:scattered_wave_energy_growth_saturate}.

When the energy density of the incident wave satisfies $\varepsilon_{\mathrm{inci}}<\varepsilon_{\mathrm{scat,max}}$, induced Compton scattering may not saturate. In this case, most of the energy of the incident wave can be transferred to the scattered wave, and the incident wave can be strongly attenuated. We refer to this regime as \emph{full scattering}. The detailed nonlinear evolution in the full scattering regime will be examined in a companion paper (Nishiura {\it et al.} in prep.).

The boundary between partial and full scattering is defined by the condition that the energy density of the incident wave reaches the saturation value in Eq. \eqref{eq:scattered_wave_energy_growth_saturate},
\begin{equation}
\varepsilon_{\mathrm{inci}}=\frac{|\bm{E}_{\mathrm{inci}}|^{2}
             + |\bm{B}_{\mathrm{inci}}|^{2}}{8\pi}\sim\varepsilon_{\mathrm{scat,max}}.
\label{eq:saturation_condition}
\end{equation}
Using Eqs.~\eqref{eq:sigma}, \eqref{eq:eta}, and \eqref{eq:scattered_wave_energy_growth_saturate}, we obtain
\begin{equation}
\begin{aligned}
\eta_{\mathrm{inci}}^{2}\sigma
&=\frac{1}{2}\frac{v_{\mathrm{th}}v_{\mathrm{A}}}{c^{2}}
\left(1+\cfrac{v_{\mathrm{A}}^{2}}{c^{2}}\right)^{-1},\\
&\sim \frac{v_{\mathrm{th}}}{4c}.
\qquad (v_{\mathrm{A}}\sim c)
\end{aligned}
\end{equation}
Therefore we expect that the nonlinear evolution can be summarized as
\begin{equation}
\begin{aligned}
\eta_{\mathrm{inci}}^{2}\,\sigma &> \frac{v_{\mathrm{th}}}{4c}
&&\rightarrow \text{partial scattering},\\
\eta_{\mathrm{inci}}^{2}\,\sigma &\le \frac{v_{\mathrm{th}}}{4c}
&&\rightarrow \text{full scattering}.
\end{aligned}
\end{equation}

\section{Discussion and Conclusion} \label{sec:summary}
In this study, we investigated the charged and neutral modes of induced Compton scattering in a magnetized electron-positron pair plasma using PIC simulations.
We demonstrated that the maximum linear growth rates, growth wavenumber, and qualitative features of density fluctuations obtained from the simulations are in good agreement with theoretical predictions. 
The charged and neutral modes exhibit distinct characteristics in terms of density fluctuations.  
In the charged mode simulations (Runs~1--15), as expected, the difference of electron and positron fluctuations $(|\widetilde{\delta n}_+ - \widetilde{\delta n}_-| / n_0)^2$ (Langmuir-like density perturbation) dominates.
Conversely, in the neutral mode simulations (Runs~16--20), the sum $(|\widetilde{\delta n}_+ + \widetilde{\delta n}_-| / n_0)^2$ (acoustic-like density fluctuation) dominates, again consistent with theoretical expectations.
The linear growth rate also shows good agreement between the simulation results and the analytical estimate, including the dependence on the physical parameters.
These results confirm that both the charged and neutral modes predicted for induced Compton scattering in a magnetized electron-positron pair plasma are successfully reproduced in the simulations.

Parametric instabilities generally compete with other processes, depending on the plasma composition and physical parameters. 
In particular, there can be a transition between induced Compton scattering and stimulated Brillouin and Raman scattering \cite{edwards16, schluck17, nishiura25b}.
For FRB emission originating from magnetars, the dominant process is expected to change as the emission propagates outward from the magnetar magnetosphere into the wind region. 
Recently, we presented a unified theoretical framework for induced scattering--including induced Compton, stimulated Brillouin, and stimulated Raman scattering--in strongly magnetized electron-positron pair plasmas \cite{nishiura25b}. 
We are currently numerically verifying these analytical results using the same methodology as in the present paper and applying them to FRBs (Nishiura {\it et al.} in prep.).

This work suggests that, even if the conditions for linear growth of induced scattering are satisfied, saturation may occur, and scattering might not actually take place.
Hence, due to saturation, FRB emission could escape from a magnetized electron-positron pair plasma without significant energy loss.
Understanding saturation could lead to constraints on the FRB emission region and the emission mechanism operating there.

In this study, we have focused on the regime where the relative amplitude $\eta_{\rm inci}^{\rm circ}$ is less than unity.
In a magnetar magnetosphere, however, the dipolar background magnetic
field decreases dramatically as waves propagate outward.
As a consequence, the relative wave amplitude can exceed unity, and strongly nonlinear wave behavior is expected \cite{chen22,beloborodov23,beloborodov24,vanthieghem25}.
As future work, we will investigate parametric instabilities in the regime $\eta_{\rm inci}^{\rm circ} > 1$.

In this work, we performed one-dimensional PIC simulations.
As a result, mode conversion processes involving waves propagating obliquely or perpendicularly to the background magnetic field (e.g. mode conversion from two Alfv\'en waves to a fast magnetosonic wave  \cite{thompson98,golbraikh23}) are not included.
As future work, we conduct two-dimensional simulations to investigate parametric instabilities and mode conversion processes involving modes that propagate across the background magnetic field.

\begin{acknowledgments}
We thank Wataru Ishizaki for valuable comments. 
Numerical computations were carried out on Cray XC50 and XD2000 at Center for Computational Astrophysics, National Astronomical Observatory of Japan, Yukawa-21 at YITP in Kyoto University, and Flow in Nagoya University through the HPCI System Research Project (Project ID: hp240147, hp250036).
S.K., K.I., and M.I. are supported by MEXT/JSPS KAKENHI Grant No.22H00130.
S.K. and M.I. are supported by the MEXT/JSPS KAKENHI Grant No.23K20038.
R.N. is supported by JST SPRING, Grant No. JPMJSP2110, and JSPS KAKENHI, Grant No. 25KJ1562. 
K.I. is supported by MEXT/JSPS KAKENHI Grant No.23H01172, 23H05430, 23H04900, 22H00130.
The authors thank the Yukawa Institute for Theoretical Physics at Kyoto University, where this work was further developed during the YITP-W-25-08 on "Exploring Extreme Transients: Frontiers in the Early Universe and Time-Domain Astronomy".
\end{acknowledgments}

\appendix
\section{Numerical Convergence} \label{sec:nc}
\begin{figure*}[htbp]
    \centering
    \includegraphics[width=1.0\linewidth]{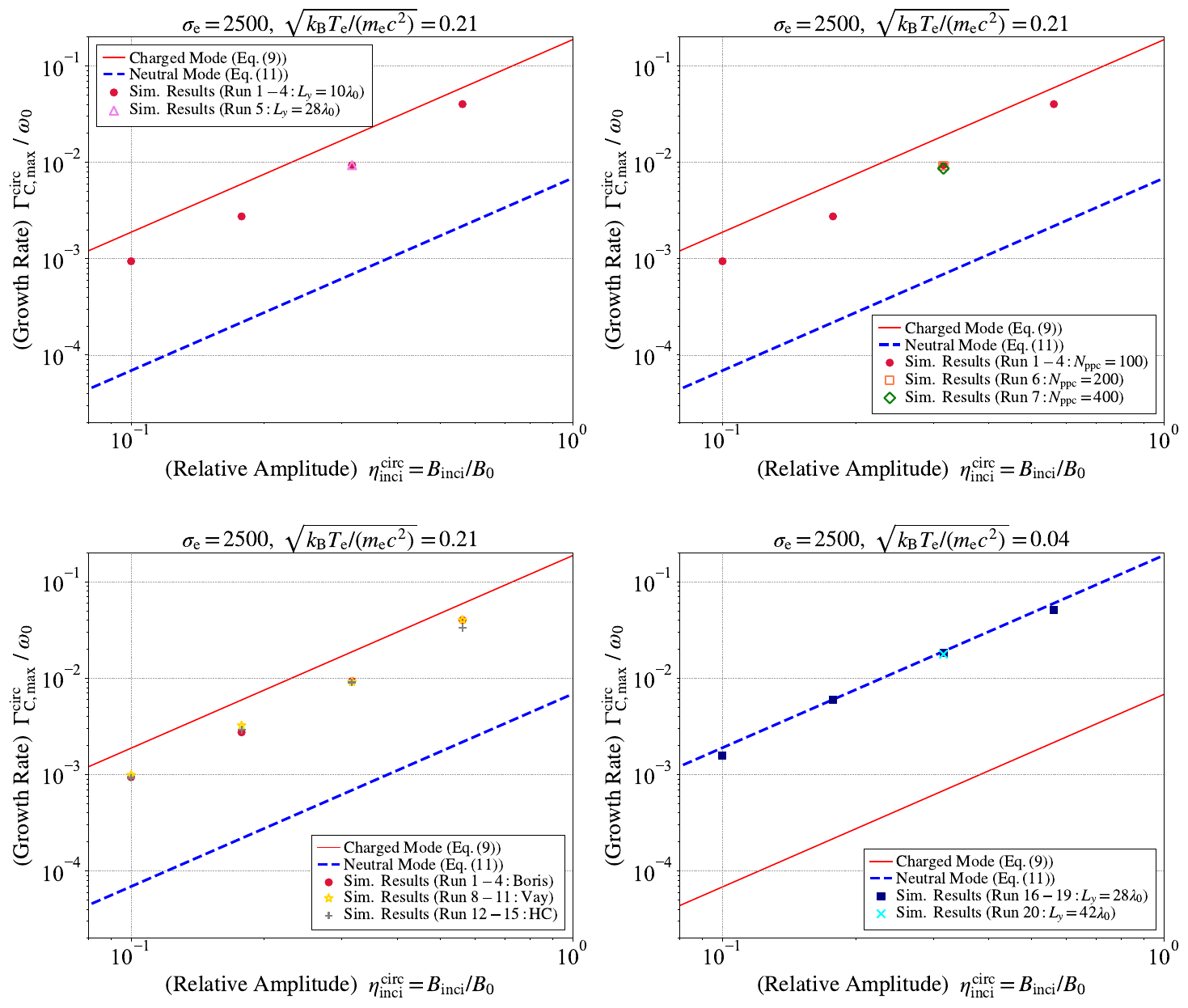}
    \phantomsection
    \caption{\RaggedRight 
    Comparison of the linear growth rate $\Gamma_{\rm C,max}^{\rm circ}$ for different box sizes $L_y$, the number of particles per cell $n_{\rm ppc}$, and particle pushers.  
    The vertical axis shows the growth rate $\Gamma_{\rm C,max}^{\rm circ}$ normalized by the incident wave frequency $\omega_0$.
    The horizontal axis is the initial amplitude of the circularly polarized incident Alfv\'en wave $\eta^{\rm circ}_{\rm inci}$.
    The red solid and blue dashed lines represent the analytical linear growth rates of the charged mode, $\Gamma_{\rm C,max}^{\rm circ,charged}$ (Eq.~(\ref{eq:gr_circ_c})) and the neutral mode, $\Gamma_{\rm C,max}^{\rm circ,neutral}$ (Eq.~(\ref{eq:gr_circ_n})), respectively.  
    The top two and bottom-left panels correspond to $\sqrt{k_{\rm B}T_{\rm e}/(m_{\rm e}c^2)} = 0.21$, where the charged mode dominates.  
    In these panels, the filled red circles indicate the results of fiducial simulations (Runs~1--4: $L_y = 10\lambda_0$, $n_{\rm ppc}=100$, Boris method \protect\cite{birdsall91}).  
    The open violet triangle in the top-left panel shows the result for a larger box (Run~5: $L_y = 28\lambda_0$).  
    The open orange square and green diamond in the top-right panel show the results for larger numbers of particles per cell (Run~6: $n_{\rm ppc}=200$ and Run~7: $n_{\rm ppc}=400$).  
    The open yellow stars and gray plus symbols in the bottom-left panel show results obtained using different particle pushers (Runs~8--11: Vay method \protect\cite{vay08} and Runs~12--15: Higuera-Cary (HC) method \protect\cite{higuera17}).  
    The bottom-right panel corresponds to $\sqrt{k_{\rm B}T_{\rm e}/(m_{\rm e}c^2)} = 0.04$, where the neutral mode dominates.  
    In this panel, the filled blue squares indicate the fiducial simulations (Runs~16--19: $L_y = 28\lambda_0$, $n_{\rm ppc}=100$, Boris method), and the cyan cross shows the result for a larger box (Run~20: $L_y = 42\lambda_0$).
    }
    \label{fig:gmax_conv}
\end{figure*}
In this section, we present a numerical convergence of the linear growth rates for the charged and neutral modes. 
For the charged mode, we examine the dependence of the results on the box size along the background magnetic field $L_y$, the number of particles per cell $n_{\rm ppc}$, and the choice of particle pusher. 
For the neutral mode, 
we only investigate the dependence on $L_y$.
The maximum growth rates obtained from simulations $\Gamma_{\rm C,max}^{\rm circ,sim}$ are summarized in Table~\ref{tab:param}.

Figure~\ref{fig:gmax_conv} shows the linear growth rate $\Gamma_{\rm C,max}^{\rm circ}$ for different values of the box size $L_y$, the number of particles per cell $n_{\rm ppc}$, and the particle pusher.  
Similar to Fig.~\ref{fig:gmax}, all panels display the maximum growth rate as a function of the incident wave amplitude $\eta_{\rm inci}^{\rm circ}$.  
The red solid and blue dashed lines indicate the analytical linear growth rates of the charged mode $\Gamma_{\rm C,max}^{\rm circ,charged}$ (Eq.~(\ref{eq:gr_circ_c})) and the neutral mode $\Gamma_{\rm C,max}^{\rm circ,neutral}$ (Eq.~(\ref{eq:gr_circ_n})), respectively.  
The top two panels and the bottom-left panel correspond to $\sqrt{k_{\rm B}T_{\rm e}/(m_{\rm e}c^2)}=0.21$, where the charged mode dominates.  
In these panels, the filled red circles represent the results of the fiducial simulations 
(Runs~1--4: $L_y = 10\lambda_0$, $n_{\rm ppc} = 100$, Boris method \cite{birdsall91}).

We first consider the case in which the charged mode dominates and the simulation box size $L_y$ is varied.  
The box size determines the resolution in wavenumber space when the electromagnetic fields are Fourier-transformed.  
If $L_y$ is too small, the wavenumber resolution of the scattered wave becomes insufficient, making it difficult to resolve individual modes.
This leads to an underestimated growth rate.  
The top-left panel of Fig.~\ref{fig:gmax_conv} compares the linear growth rates 
for different box sizes.  
The open violet triangle denotes the result for a larger box size (Run~5: $L_y = 28\lambda_0$).  
As shown, the maximum growth rate obtained from the larger box size simulation is in good agreement with that from the fiducial box size.  
Based on this comparison, we adopt $L_y = 10\lambda_0$ as the fiducial value for the charged mode in this study.

Next, we consider the case in which the charged mode dominates and the number of particles per cell $n_{\rm ppc}$ is varied.  
The top-right panel of Fig.~\ref{fig:gmax_conv} compares the linear growth rates 
for different values of $n_{\rm ppc}$.  
The open orange square and green diamond indicate the results for larger values of $n_{\rm ppc}$ 
(Run~6: $n_{\rm ppc} = 200$ and Run~7: $n_{\rm ppc} = 400$).  
As shown in the top-right panel of Fig.~\ref{fig:gmax_conv}, the maximum growth rates obtained from the larger $n_{\rm ppc}$ simulations are in good agreement with those from the fiducial case. 
Based on this comparison, we adopt $n_{\rm ppc} = 100$ as the fiducial value for the simulations in this study.

Next, we consider the case in which the charged mode dominates and the particle pusher is varied.  
The bottom-left panel of Fig.~\ref{fig:gmax_conv} compares the linear growth rates for different particle pushers.  
The open yellow stars and gray plus symbols correspond to the results obtained with other pushers 
(Runs~8--11: Vay method \cite{vay08} and Runs~12--15: Higuera-Cary (HC) method \cite{higuera17}).  
As shown, the maximum growth rates from all simulations are in good agreement.  
Accordingly, we adopt the Boris method as the fiducial particle pusher in this study.

Finally, we consider the case in which the neutral mode dominates and the box size $L_y$ is varied.  
The bottom-right panel of Fig.~\ref{fig:gmax_conv} corresponds to $\sqrt{k_{\rm B}T_{\rm e}/(m_{\rm e}c^2)}=0.04$ and shows the linear growth rates for different $L_y$ values.  
The filled blue squares represent the results for the fiducial box size 
(Runs~16--19: $L_y = 28\lambda_0$), while the cyan cross corresponds to the result 
for a larger box (Run~20: $L_y = 42\lambda_0$).  
As in the charged mode case, the maximum growth rate obtained from the larger 
box size simulation is in good agreement with that from the fiducial simulation.  
Therefore, we adopt $L_y = 28\lambda_0$ as the fiducial box size for the neutral mode 
simulations.

\FloatBarrier
\nocite{*}

\bibliographystyle{apsrev4-2}
\bibliography{reference}

@PREAMBLE{
 "\providecommand{\noopsort}[1]{}" 
 # "\providecommand{\singleletter}[1]{#1}%" 
}

@ARTICLE{esirkepov01,
       author = {{Esirkepov}, T. Zh.},
        title = "{Exact charge conservation scheme for Particle-in-Cell simulation with an arbitrary form-factor}",
      journal = {Computer Physics Communications},
         year = 2001,
        month = apr,
       volume = {135},
       number = {2},
        pages = {144-153},
          doi = {10.1016/S0010-4655(00)00228-9},
       adsurl = {https://ui.adsabs.harvard.edu/abs/2001CoPhC.135..144E}
}

@ARTICLE{nishiura25a,
       author = {{Nishiura}, Rei and {Kamijima}, Shoma F. and {Iwamoto}, Masanori and {Ioka}, Kunihito},
        title = "{Induced Compton scattering in magnetized electron and positron pair plasma}",
      journal = {\prd},
         year = 2025,
        month = mar,
       volume = {111},
       number = {6},
          eid = {063055},
        pages = {063055},
          doi = {10.1103/PhysRevD.111.063055},
archivePrefix = {arXiv},
       eprint = {2411.00936},
 primaryClass = {astro-ph.HE},
       adsurl = {https://ui.adsabs.harvard.edu/abs/2025PhRvD.111f3055N}
}

@software{matsumoto24,
       author = {{Matsumoto}, Yosuke and {Amano}, Takanobu and {Iwamoto}, Masanori and {Jikei}, Taiki and {Zenitani}, Seiji},
        title = "{WumingCode/WumingPIC2D: Version 0.6}",
         year = 2024,
        month = aug,
          eid = {10.5281/zenodo.13269434},
          doi = {10.5281/zenodo.13269434},
      version = {v0.6},
    publisher = {Zenodo},
       adsurl = {https://ui.adsabs.harvard.edu/abs/2024zndo..10990576A}
}

@ARTICLE{ikeya15,
       author = {{Ikeya}, Naoki and {Matsumoto}, Yosuke},
        title = "{Stability property of numerical Cherenkov radiation and its application to relativistic shock simulations}",
      journal = {\pasj},
         year = 2015,
        month = aug,
       volume = {67},
       number = {4},
          eid = {64},
        pages = {64},
          doi = {10.1093/pasj/psv052},
archivePrefix = {arXiv},
       eprint = {1412.2480},
 primaryClass = {astro-ph.HE},
       adsurl = {https://ui.adsabs.harvard.edu/abs/2015PASJ...67...64I}
}

@ARTICLE{zenitani24,
       author = {{Zenitani}, Seiji},
        title = "{Modifications to Swisdak (2013)'s rejection sampling algorithm for a Maxwell{\textendash}J{\"u}ttner distribution in particle simulations}",
      journal = {Physics of Plasmas},
         year = 2024,
        month = sep,
       volume = {31},
       number = {9},
          eid = {094501},
        pages = {094501},
          doi = {10.1063/5.0226859},
archivePrefix = {arXiv},
       eprint = {2408.09105},
 primaryClass = {physics.plasm-ph},
       adsurl = {https://ui.adsabs.harvard.edu/abs/2024PhPl...31i4501Z}
}

@ARTICLE{matsukiyo03,
       author = {{Matsukiyo}, S. and {Hada}, T.},
        title = "{Parametric instabilities of circularly polarized Alfv{\'e}n waves in a relativistic electron-positron plasma}",
      journal = {\pre},
         year = 2003,
        month = apr,
       volume = {67},
       number = {4},
          eid = {046406},
        pages = {046406},
          doi = {10.1103/PhysRevE.67.046406},
       adsurl = {https://ui.adsabs.harvard.edu/abs/2003PhRvE..67d6406M}
}

@ARTICLE{terasawa86,
       author = {{Terasawa}, T. and {Hoshino}, M. and {Sakai}, J. -I. and {Hada}, T.},
        title = "{Decay instability of finite-amplitude circularly polarized Alfven waves: A numerical simulation of stimulated Brillouin scattering}",
      journal = {\jgr},
         year = 1986,
        month = apr,
       volume = {91},
       number = {A4},
        pages = {4171-4187},
          doi = {10.1029/JA091iA04p04171},
       adsurl = {https://ui.adsabs.harvard.edu/abs/1986JGR....91.4171T}
}

@ARTICLE{amano13,
       author = {{Amano}, Takanobu and {Kirk}, John G.},
        title = "{The Role of Superluminal Electromagnetic Waves in Pulsar Wind Termination Shocks}",
      journal = {\apj},
         year = 2013,
        month = jun,
       volume = {770},
       number = {1},
          eid = {18},
        pages = {18},
          doi = {10.1088/0004-637X/770/1/18},
archivePrefix = {arXiv},
       eprint = {1303.2702},
 primaryClass = {astro-ph.HE},
       adsurl = {https://ui.adsabs.harvard.edu/abs/2013ApJ...770...18A}
}

@ARTICLE{schluck17,
       author = {{Schluck}, F. and {Lehmann}, G. and {Spatschek}, K.~H.},
        title = "{Parametric pulse amplification by acoustic quasimodes in electron-positron plasma}",
      journal = {\pre},
         year = 2017,
        month = nov,
       volume = {96},
       number = {5},
          eid = {053204},
        pages = {053204},
          doi = {10.1103/PhysRevE.96.053204},
       adsurl = {https://ui.adsabs.harvard.edu/abs/2017PhRvE..96e3204S}
}

@ARTICLE{edwards16,
       author = {{Edwards}, Matthew R. and {Fisch}, Nathaniel J. and {Mikhailova}, Julia M.},
        title = "{Strongly Enhanced Stimulated Brillouin Backscattering in an Electron-Positron Plasma}",
      journal = {\prl},
         year = 2016,
        month = jan,
       volume = {116},
       number = {1},
          eid = {015004},
        pages = {015004},
          doi = {10.1103/PhysRevLett.116.015004},
archivePrefix = {arXiv},
       eprint = {1512.00744},
 primaryClass = {physics.plasm-ph},
       adsurl = {https://ui.adsabs.harvard.edu/abs/2016PhRvL.116a5004E}
}

@ARTICLE{ghosh22,
       author = {{Ghosh}, Arka and {Kagan}, Daniel and {Keshet}, Uri and {Lyubarsky}, Yuri},
        title = "{Nonlinear Electromagnetic-wave Interactions in Pair Plasma. I. Nonrelativistic Regime}",
      journal = {\apj},
         year = 2022,
        month = may,
       volume = {930},
       number = {2},
          eid = {106},
        pages = {106},
          doi = {10.3847/1538-4357/ac581d},
archivePrefix = {arXiv},
       eprint = {2111.00656},
 primaryClass = {astro-ph.HE},
       adsurl = {https://ui.adsabs.harvard.edu/abs/2022ApJ...930..106G}
}

@ARTICLE{zenitani15,
       author = {{Zenitani}, Seiji},
        title = "{Loading relativistic Maxwell distributions in particle simulations}",
      journal = {Physics of Plasmas},
         year = 2015,
        month = apr,
       volume = {22},
       number = {4},
          eid = {042116},
        pages = {042116},
          doi = {10.1063/1.4919383},
archivePrefix = {arXiv},
       eprint = {1504.03910},
 primaryClass = {astro-ph.HE},
       adsurl = {https://ui.adsabs.harvard.edu/abs/2015PhPl...22d2116Z}
}

@ARTICLE{vay08,
       author = {{Vay}, J. -L.},
        title = "{Simulation of beams or plasmas crossing at relativistic velocity}",
      journal = {Physics of Plasmas},
         year = 2008,
        month = may,
       volume = {15},
       number = {5},
          eid = {056701},
        pages = {056701},
          doi = {10.1063/1.2837054},
       adsurl = {https://ui.adsabs.harvard.edu/abs/2008PhPl...15e6701V}
}

@ARTICLE{higuera17,
       author = {{Higuera}, A.~V. and {Cary}, J.~R.},
        title = "{Structure-preserving second-order integration of relativistic charged particle trajectories in electromagnetic fields}",
      journal = {Physics of Plasmas},
         year = 2017,
        month = may,
       volume = {24},
       number = {5},
          eid = {052104},
        pages = {052104},
          doi = {10.1063/1.4979989},
archivePrefix = {arXiv},
       eprint = {1701.05605},
 primaryClass = {physics.plasm-ph},
       adsurl = {https://ui.adsabs.harvard.edu/abs/2017PhPl...24e2104H}
}

@BOOK{birdsall91,
       author = {{Birdsall}, C.~K. and {Langdon}, A.~B.},
        title = "{Plasma Physics via Computer Simulation}",
         year = 1991,
       adsurl = {https://ui.adsabs.harvard.edu/abs/1991ppcs.book.....B}
}

@ARTICLE{drake74,
       author = {{Drake}, J.~F. and {Kaw}, P.~K. and {Lee}, Y.~C. and {Schmid}, G. and {Liu}, C.~S. and {Rosenbluth}, Marshall N.},
        title = "{Parametric instabilities of electromagnetic waves in plasmas}",
      journal = {Physics of Fluids},
         year = 1974,
        month = apr,
       volume = {17},
       number = {4},
        pages = {778-785},
          doi = {10.1063/1.1694789},
       adsurl = {https://ui.adsabs.harvard.edu/abs/1974PhFl...17..778D}
}

@ARTICLE{forslund75,
       author = {{Forslund}, D.~W. and {Kindel}, J.~M. and {Lindman}, E.~L.},
        title = "{Theory of stimulated scattering processes in laser-irradiated plasmas}",
      journal = {Physics of Fluids},
         year = 1975,
        month = aug,
       volume = {18},
       number = {8},
        pages = {1002-1016},
          doi = {10.1063/1.861248},
       adsurl = {https://ui.adsabs.harvard.edu/abs/1975PhFl...18.1002F}
}

@ARTICLE{cohen79,
       author = {{Cohen}, B.~I. and {Max}, C.~E.},
        title = "{Stimulated scattering of light by ion modes in a homogeneous plasma: Space-time evolution}",
      journal = {Physics of Fluids},
         year = 1979,
        month = jun,
       volume = {22},
       number = {6},
        pages = {1115-1132},
          doi = {10.1063/1.862713},
       adsurl = {https://ui.adsabs.harvard.edu/abs/1979PhFl...22.1115C}
}

@ARTICLE{kaw73,
       author = {{Kaw}, P. and {Schmidt}, G. and {Wilcox}, T.},
        title = "{Filamentation and trapping of electromagnetic radiation in plasmas}",
      journal = {Physics of Fluids},
         year = 1973,
        month = sep,
       volume = {16},
       number = {9},
        pages = {1522-1525},
          doi = {10.1063/1.1694552},
       adsurl = {https://ui.adsabs.harvard.edu/abs/1973PhFl...16.1522K}
}

@ARTICLE{max73,
       author = {{Max}, C.~E.},
        title = "{Parametric instability of a relativistically strong electromagnetic wave.}",
      journal = {Physics of Fluids},
         year = 1973,
        month = jan,
       volume = {16},
        pages = {1480-1489},
          doi = {10.1063/1.1694545},
       adsurl = {https://ui.adsabs.harvard.edu/abs/1973PhFl...16.1480M}
}

@ARTICLE{max74,
       author = {{Max}, Claire Ellen and {Arons}, Jonathan and {Langdon}, A. Bruce},
        title = "{Self-Modulation and Self-Focusing of Electromagnetic Waves in Plasmas}",
      journal = {\prl},
         year = 1974,
        month = jul,
       volume = {33},
       number = {4},
        pages = {209-212},
          doi = {10.1103/PhysRevLett.33.209},
       adsurl = {https://ui.adsabs.harvard.edu/abs/1974PhRvL..33..209M}
}

@PROCEEDINGS{kruer88,
        title = "{The physics of laser plasma interactions}",
    booktitle = {The physics of laser plasma interactions},
         year = 1988,
       editor = {{Kruer}, William L.},
       volume = {73},
        month = jan,
       adsurl = {https://ui.adsabs.harvard.edu/abs/1988aw.....73.....K}
}

@ARTICLE{mima75,
       author = {{Mima}, K. and {Nishikawa}, K.},
        title = "{Theory of Cascade Induced Forward Raman Scattering in a Plasma}",
      journal = {Journal of the Physical Society of Japan},
         year = 1975,
        month = jun,
       volume = {38},
       number = {6},
        pages = {1742-1752},
          doi = {10.1143/JPSJ.38.1742},
       adsurl = {https://ui.adsabs.harvard.edu/abs/1975JPSJ...38.1742M}
}

@BOOK{sagdeev69,
       author = {{Sagdeev}, R.~Z. and {Galeev}, A.~A.},
        title = "{Nonlinear Plasma Theory}",
         year = 1969,
       adsurl = {https://ui.adsabs.harvard.edu/abs/1969npt..book.....S}
}

@ARTICLE{galeev63,
       author = {{Galeev}, A.~A. and {Oraevskii}, V.~N.},
        title = "{The Stability of Alfv{\'e}n Waves}",
      journal = {Soviet Physics Doklady},
         year = 1963,
        month = may,
       volume = {7},
        pages = {988},
       adsurl = {https://ui.adsabs.harvard.edu/abs/1963SPhD....7..988G}
}

@ARTICLE{barnes66,
       author = {{Barnes}, Aaron},
        title = "{Collisionless Damping of Hydromagnetic Waves}",
      journal = {Physics of Fluids},
         year = 1966,
        month = aug,
       volume = {9},
       number = {8},
        pages = {1483-1495},
          doi = {10.1063/1.1761882},
       adsurl = {https://ui.adsabs.harvard.edu/abs/1966PhFl....9.1483B}
}

@ARTICLE{derby78,
       author = {{Derby}, Jr., N.~F.},
        title = "{Modulational instability of finite-amplitude, circularly polarized Alfv{\'e}n waves.}",
      journal = {\apj},
         year = 1978,
        month = sep,
       volume = {224},
        pages = {1013-1016},
          doi = {10.1086/156451},
       adsurl = {https://ui.adsabs.harvard.edu/abs/1978ApJ...224.1013D}
}

@ARTICLE{inhester90,
       author = {{Inhester}, B.},
        title = "{A drift-kinetic treatment of the parametric decay of large-amplitude Alfven waves}",
      journal = {\jgr},
         year = 1990,
        month = jul,
       volume = {95},
        pages = {10525-10539},
          doi = {10.1029/JA095iA07p10525},
       adsurl = {https://ui.adsabs.harvard.edu/abs/1990JGR....9510525I}
}

@ARTICLE{jayanti93,
       author = {{Jayanti}, Venku and {Hollweg}, Joseph V.},
        title = "{On the dispersion relations for parametric instabilities of parallel-progagating Alfv{\'e}n waves}",
      journal = {\jgr},
         year = 1993,
        month = aug,
       volume = {98},
       number = {A8},
        pages = {13247-13252},
          doi = {10.1029/93JA00920},
       adsurl = {https://ui.adsabs.harvard.edu/abs/1993JGR....9813247J}
}

@ARTICLE{goldstein78,
       author = {{Goldstein}, M.~L.},
        title = "{An instability of finite amplitude circularly polarized Afv{\'e}n waves.}",
      journal = {\apj},
         year = 1978,
        month = jan,
       volume = {219},
        pages = {700-704},
          doi = {10.1086/155829},
       adsurl = {https://ui.adsabs.harvard.edu/abs/1978ApJ...219..700G}
}

@ARTICLE{hoshino89,
       author = {{Hoshino}, M. and {Goldstein}, M.~L.},
        title = "{Time evolution from linear to nonlinear stages in magnetohydrodynamic parametric instabilities}",
      journal = {Physics of Fluids B},
         year = 1989,
        month = jul,
       volume = {1},
       number = {7},
        pages = {1405-1415},
          doi = {10.1063/1.858971},
       adsurl = {https://ui.adsabs.harvard.edu/abs/1989PhFlB...1.1405H}
}

@ARTICLE{hollweg94,
       author = {{Hollweg}, Joseph V.},
        title = "{Beat, modulational, and decay instabilities of a circularly polarized Alfv{\'e}n wave}",
      journal = {\jgr},
         year = 1994,
        month = dec,
       volume = {99},
       number = {A12},
        pages = {23431-23448},
          doi = {10.1029/94JA02185},
       adsurl = {https://ui.adsabs.harvard.edu/abs/1994JGR....9923431H}
}

@ARTICLE{longtin86,
       author = {{Longtin}, M. and {Sonnerup}, B.~U.~O.},
        title = "{Modulation instability of circularly polarized Alfv{\'e}n waves}",
      journal = {\jgr},
         year = 1986,
        month = jun,
       volume = {91},
       number = {A6},
        pages = {6816-6824},
          doi = {10.1029/JA091iA06p06816},
       adsurl = {https://ui.adsabs.harvard.edu/abs/1986JGR....91.6816L}
}

@ARTICLE{delzanna01,
       author = {{Del Zanna}, L. and {Velli}, M. and {Londrillo}, P.},
        title = "{Parametric decay of circularly polarized Alfv{\'e}n waves: Multidimensional simulations in periodic and open domains}",
      journal = {\aap},
         year = 2001,
        month = feb,
       volume = {367},
        pages = {705-718},
          doi = {10.1051/0004-6361:20000455},
       adsurl = {https://ui.adsabs.harvard.edu/abs/2001A&A...367..705D}
}

@ARTICLE{nariyuki06,
       author = {{Nariyuki}, Y. and {Hada}, T.},
        title = "{Kinetically modified parametric instabilities of circularly polarized Alfv{\'e}n waves: Ion kinetic effects}",
      journal = {Physics of Plasmas},
         year = 2006,
        month = dec,
       volume = {13},
       number = {12},
          eid = {124501},
        pages = {124501},
          doi = {10.1063/1.2399468},
archivePrefix = {arXiv},
       eprint = {physics/0608306},
 primaryClass = {physics.plasm-ph},
       adsurl = {https://ui.adsabs.harvard.edu/abs/2006PhPl...13l4501N}
}

@ARTICLE{delzenna15,
       author = {{Del Zanna}, L. and {Matteini}, L. and {Landi}, S. and {Verdini}, A. and {Velli}, M.},
        title = "{Parametric decay of parallel and oblique Alfv{\'e}n waves in the expanding solar wind}",
      journal = {Journal of Plasma Physics},
         year = 2015,
        month = jan,
       volume = {81},
       number = {1},
          eid = {325810102},
        pages = {325810102},
          doi = {10.1017/S0022377814000579},
archivePrefix = {arXiv},
       eprint = {1407.5851},
 primaryClass = {astro-ph.SR},
       adsurl = {https://ui.adsabs.harvard.edu/abs/2015JPlPh..81a3202D}
}

@ARTICLE{nariyuki08,
       author = {{Nariyuki}, Y. and {Matsukiyo}, S. and {Hada}, T.},
        title = "{Parametric instabilities of large-amplitude parallel propagating Alfv{\'e}n waves: 2D PIC simulation}",
      journal = {New Journal of Physics},
         year = 2008,
        month = aug,
       volume = {10},
       number = {8},
          eid = {083004},
        pages = {083004},
          doi = {10.1088/1367-2630/10/8/083004},
archivePrefix = {arXiv},
       eprint = {0804.4041},
 primaryClass = {astro-ph},
       adsurl = {https://ui.adsabs.harvard.edu/abs/2008NJPh...10h3004N}
}

@ARTICLE{shoda16,
       author = {{Shoda}, M. and {Yokoyama}, T.},
        title = "{Nonlinear Reflection Process of Linearly Polarized, Broadband Alfv{\'e}n Waves in the Fast Solar Wind}",
      journal = {\apj},
         year = 2016,
        month = apr,
       volume = {820},
       number = {2},
          eid = {123},
        pages = {123},
          doi = {10.3847/0004-637X/820/2/123},
archivePrefix = {arXiv},
       eprint = {1602.03628},
 primaryClass = {astro-ph.SR},
       adsurl = {https://ui.adsabs.harvard.edu/abs/2016ApJ...820..123S}
}

@ARTICLE{suzuki05,
       author = {{Suzuki}, Takeru K. and {Inutsuka}, Shu-ichiro},
        title = "{Making the Corona and the Fast Solar Wind: A Self-consistent Simulation for the Low-Frequency Alfv{\'e}n Waves from the Photosphere to 0.3 AU}",
      journal = {\apjl},
         year = 2005,
        month = oct,
       volume = {632},
       number = {1},
        pages = {L49-L52},
          doi = {10.1086/497536},
archivePrefix = {arXiv},
       eprint = {astro-ph/0506639},
 primaryClass = {astro-ph},
       adsurl = {https://ui.adsabs.harvard.edu/abs/2005ApJ...632L..49S}
}

@ARTICLE{suzuki06,
       author = {{Suzuki}, Takeru K. and {Inutsuka}, Shu-Ichiro},
        title = "{Solar winds driven by nonlinear low-frequency Alfv{\'e}n waves from the photosphere: Parametric study for fast/slow winds and disappearance of solar winds}",
      journal = {Journal of Geophysical Research (Space Physics)},
         year = 2006,
        month = jun,
       volume = {111},
       number = {A6},
          eid = {A06101},
        pages = {A06101},
          doi = {10.1029/2005JA011502},
archivePrefix = {arXiv},
       eprint = {astro-ph/0511006},
 primaryClass = {astro-ph},
       adsurl = {https://ui.adsabs.harvard.edu/abs/2006JGRA..111.6101S}
}

@ARTICLE{shoda18,
       author = {{Shoda}, Munehito and {Yokoyama}, Takaaki and {Suzuki}, Takeru K.},
        title = "{Frequency-dependent Alfv{\'e}n-wave Propagation in the Solar Wind: Onset and Suppression of Parametric Decay Instability}",
      journal = {\apj},
         year = 2018,
        month = jun,
       volume = {860},
       number = {1},
          eid = {17},
        pages = {17},
          doi = {10.3847/1538-4357/aac218},
archivePrefix = {arXiv},
       eprint = {1803.02606},
 primaryClass = {astro-ph.SR},
       adsurl = {https://ui.adsabs.harvard.edu/abs/2018ApJ...860...17S}
}

@ARTICLE{shi17,
       author = {{Shi}, Mijie and {Li}, Hui and {Xiao}, Chijie and {Wang}, Xiaogang},
        title = "{The Parametric Decay Instability of Alfv{\'e}n Waves in Turbulent Plasmas and the Applications in the Solar Wind}",
      journal = {\apj},
         year = 2017,
        month = jun,
       volume = {842},
       number = {1},
          eid = {63},
        pages = {63},
          doi = {10.3847/1538-4357/aa71b6},
archivePrefix = {arXiv},
       eprint = {1705.03829},
 primaryClass = {physics.space-ph},
       adsurl = {https://ui.adsabs.harvard.edu/abs/2017ApJ...842...63S}
}

@ARTICLE{nariyuki22,
       author = {{Nariyuki}, Yasuhiro},
        title = "{Low-frequency Alfv{\'e}n waves and parametric instabilities in fluid and kinetic plasmas}",
      journal = {Reviews of Modern Plasma Physics},
         year = 2022,
        month = dec,
       volume = {6},
       number = {1},
          eid = {22},
        pages = {22},
          doi = {10.1007/s41614-022-00085-1},
       adsurl = {https://ui.adsabs.harvard.edu/abs/2022RvMPP...6...22N}
}

@ARTICLE{thide82,
       author = {{Thid{\'e}}, Bo and {Kopka}, Helmut and {Stubbe}, Peter},
        title = "{Observations of stimulated scattering of a strong high-frequency radio wave in the ionosphere}",
      journal = {\prl},
         year = 1982,
        month = nov,
       volume = {49},
       number = {21},
        pages = {1561-1564},
          doi = {10.1103/PhysRevLett.49.1561},
       adsurl = {https://ui.adsabs.harvard.edu/abs/1982PhRvL..49.1561T}
}

@ARTICLE{1989PhR...179...79R,
       author = {{Robinson}, T.~R.},
        title = "{The heating of the high lattitude ionosphere by high power radio waves}",
      journal = {\physrep},
         year = 1989,
        month = jul,
       volume = {179},
       number = {2-3},
        pages = {79-209},
          doi = {10.1016/0370-1573(89)90005-7},
       adsurl = {https://ui.adsabs.harvard.edu/abs/1989PhR...179...79R}
}

@ARTICLE{robinson89,
       author = {{Robinson}, T.~R.},
        title = "{The heating of the high lattitude ionosphere by high power radio waves}",
      journal = {\physrep},
         year = 1989,
        month = jul,
       volume = {179},
       number = {2-3},
        pages = {79-209},
          doi = {10.1016/0370-1573(89)90005-7},
       adsurl = {https://ui.adsabs.harvard.edu/abs/1989PhR...179...79R}
}

@ARTICLE{syunyaev71,
       author = {{Syunyaev}, R.~A.},
        title = "{Induced Compton Scattering by Thermal Electrons and the Low-Frequency Spectrum of Radio Sources.}",
      journal = {\sovast},
         year = 1971,
        month = oct,
       volume = {15},
        pages = {190},
       adsurl = {https://ui.adsabs.harvard.edu/abs/1971SvA....15..190S}
}

@ARTICLE{coppi93,
       author = {{Coppi}, P. and {Blandford}, R.~D. and {Rees}, M.~J.},
        title = "{Anisotropic induced compton scattering - constraints on models of active galactic nuclei.}",
      journal = {\mnras},
         year = 1993,
        month = jun,
       volume = {262},
        pages = {603-618},
          doi = {10.1093/mnras/262.3.603},
       adsurl = {https://ui.adsabs.harvard.edu/abs/1993MNRAS.262..603C}
}

@ARTICLE{blandford76,
       author = {{Blandford}, R.~D. and {Scharlemann}, E.~T.},
        title = "{On the scattering and absorption of electromagnetic radiation with pulsar magnetospheres.}",
      journal = {\mnras},
         year = 1976,
        month = jan,
       volume = {174},
        pages = {59-85},
          doi = {10.1093/mnras/174.1.59},
       adsurl = {https://ui.adsabs.harvard.edu/abs/1976MNRAS.174...59B}
}

@ARTICLE{wilson78,
       author = {{Wilson}, D.~B. and {Rees}, M.~J.},
        title = "{Induced Compton scattering in pulsar winds}",
      journal = {\mnras},
         year = 1978,
        month = oct,
       volume = {185},
        pages = {297},
          doi = {10.1093/mnras/185.2.297},
       adsurl = {https://ui.adsabs.harvard.edu/abs/1978MNRAS.185..297W}
}

@ARTICLE{wilson82,
       author = {{Wilson}, D.~B.},
        title = "{Induced compton scattering in radiative transfer}",
      journal = {\mnras},
         year = 1982,
        month = sep,
       volume = {200},
        pages = {881-906},
          doi = {10.1093/mnras/200.4.881},
       adsurl = {https://ui.adsabs.harvard.edu/abs/1982MNRAS.200..881W}
}

@ARTICLE{lyubarskii96,
       author = {{Lyubarskii}, Yu. E. and {Petrova}, S.~A.},
        title = "{Stimulated scattering of radio emission in pulsar magnetospheres}",
      journal = {Astronomy Letters},
         year = 1996,
        month = may,
       volume = {22},
       number = {3},
        pages = {399-408},
       adsurl = {https://ui.adsabs.harvard.edu/abs/1996AstL...22..399L}
}

@ARTICLE{lyubarsky08,
       author = {{Lyubarsky}, Yuri},
        title = "{Induced Scattering of Short Radio Pulses}",
      journal = {\apj},
         year = 2008,
        month = aug,
       volume = {682},
       number = {2},
        pages = {1443-1449},
          doi = {10.1086/589435},
archivePrefix = {arXiv},
       eprint = {0804.2069},
 primaryClass = {astro-ph},
       adsurl = {https://ui.adsabs.harvard.edu/abs/2008ApJ...682.1443L}
}

@ARTICLE{iwamoto23,
       author = {{Iwamoto}, Masanori and {Sobacchi}, Emanuele and {Sironi}, Lorenzo},
        title = "{Kinetic simulations of the filamentation instability in pair plasmas}",
      journal = {\mnras},
         year = 2023,
        month = jun,
       volume = {522},
       number = {2},
        pages = {2133-2144},
          doi = {10.1093/mnras/stad1100},
archivePrefix = {arXiv},
       eprint = {2304.03577},
 primaryClass = {astro-ph.HE},
       adsurl = {https://ui.adsabs.harvard.edu/abs/2023MNRAS.522.2133I}
}

@ARTICLE{sobacchi23,
       author = {{Sobacchi}, Emanuele and {Lyubarsky}, Yuri and {Beloborodov}, Andrei M. and {Sironi}, Lorenzo and {Iwamoto}, Masanori},
        title = "{Saturation of the Filamentation Instability and Dispersion Measure of Fast Radio Bursts}",
      journal = {\apjl},
         year = 2023,
        month = feb,
       volume = {943},
       number = {2},
          eid = {L21},
        pages = {L21},
          doi = {10.3847/2041-8213/acb260},
archivePrefix = {arXiv},
       eprint = {2210.08754},
 primaryClass = {astro-ph.HE},
       adsurl = {https://ui.adsabs.harvard.edu/abs/2023ApJ...943L..21S}
}

@ARTICLE{ishizaki24,
       author = {{Ishizaki}, Wataru and {Ioka}, Kunihito},
        title = "{Parametric decay instability of circularly polarized Alfv{\'e}n waves in magnetically dominated plasma}",
      journal = {\pre},
         year = 2024,
        month = jul,
       volume = {110},
       number = {1},
          eid = {015205},
        pages = {015205},
          doi = {10.1103/PhysRevE.110.015205},
archivePrefix = {arXiv},
       eprint = {2404.15689},
 primaryClass = {astro-ph.HE},
       adsurl = {https://ui.adsabs.harvard.edu/abs/2024PhRvE.110a5205I}
}

@ARTICLE{komissarov25,
       author = {{Komissarov}, Serguei S.},
        title = "{Parametric instability of Alfv{\'e}n wave packets}",
      journal = {\mnras},
         year = 2025,
        month = sep,
       volume = {542},
       number = {3},
        pages = {2510-2524},
          doi = {10.1093/mnras/staf1385},
archivePrefix = {arXiv},
       eprint = {2507.10038},
 primaryClass = {astro-ph.SR},
       adsurl = {https://ui.adsabs.harvard.edu/abs/2025MNRAS.542.2510K}
}

@ARTICLE{nishiura25b,
       author = {{Nishiura}, Rei and {Kamijima}, Shoma F. and {Ioka}, Kunihito},
        title = "{Unified kinetic theory of induced scattering: Compton, Brillouin, and Raman processes in magnetized electron and positron pair plasma}",
      journal = {arXiv e-prints},
         year = 2025,
        month = oct,
          eid = {arXiv:2510.12869},
        pages = {arXiv:2510.12869},
archivePrefix = {arXiv},
       eprint = {2510.12869},
 primaryClass = {astro-ph.HE},
       adsurl = {https://ui.adsabs.harvard.edu/abs/2025arXiv251012869N}
}

@ARTICLE{vinas91,
       author = {{Vi{\~n}as}, Adolfo F. and {Goldstein}, Melvyn L.},
        title = "{Parametric instabilities of circularly polarized large-amplitude dispersive Alfv{\'e}n waves: excitation of obliquely-propagating daughter and side-band waves}",
      journal = {Journal of Plasma Physics},
         year = 1991,
        month = aug,
       volume = {46},
       number = {1},
        pages = {129-152},
          doi = {10.1017/S0022377800015993},
       adsurl = {https://ui.adsabs.harvard.edu/abs/1991JPlPh..46..129V}
}

@ARTICLE{golbraikh23,
       author = {{Golbraikh}, Ephim and {Lyubarsky}, Yuri},
        title = "{On the Escape of Low-frequency Waves from Magnetospheres of Neutron Stars}",
      journal = {\apj},
         year = 2023,
        month = nov,
       volume = {957},
       number = {2},
          eid = {102},
        pages = {102},
          doi = {10.3847/1538-4357/acfa78},
archivePrefix = {arXiv},
       eprint = {2309.09218},
 primaryClass = {astro-ph.HE},
       adsurl = {https://ui.adsabs.harvard.edu/abs/2023ApJ...957..102G}
}

@ARTICLE{lorimer07,
       author = {{Lorimer}, D.~R. and {Bailes}, M. and {McLaughlin}, M.~A. and {Narkevic}, D.~J. and {Crawford}, F.},
        title = "{A Bright Millisecond Radio Burst of Extragalactic Origin}",
      journal = {Science},
         year = 2007,
        month = nov,
       volume = {318},
       number = {5851},
        pages = {777},
          doi = {10.1126/science.1147532},
archivePrefix = {arXiv},
       eprint = {0709.4301},
 primaryClass = {astro-ph},
       adsurl = {https://ui.adsabs.harvard.edu/abs/2007Sci...318..777L}
}

@ARTICLE{petroff19,
       author = {{Petroff}, E. and {Hessels}, J.~W.~T. and {Lorimer}, D.~R.},
        title = "{Fast radio bursts}",
      journal = {\aapr},
         year = 2019,
        month = dec,
       volume = {27},
       number = {1},
          eid = {4},
        pages = {4},
          doi = {10.1007/s00159-019-0116-6},
archivePrefix = {arXiv},
       eprint = {1904.07947},
 primaryClass = {astro-ph.HE},
       adsurl = {https://ui.adsabs.harvard.edu/abs/2019A&ARv..27....4P}
}

@ARTICLE{bzhang23,
       author = {{Zhang}, Bing},
        title = "{The physics of fast radio bursts}",
      journal = {Reviews of Modern Physics},
         year = 2023,
        month = jul,
       volume = {95},
       number = {3},
          eid = {035005},
        pages = {035005},
          doi = {10.1103/RevModPhys.95.035005},
archivePrefix = {arXiv},
       eprint = {2212.03972},
 primaryClass = {astro-ph.HE},
       adsurl = {https://ui.adsabs.harvard.edu/abs/2023RvMP...95c5005Z}
}

@ARTICLE{lyubarsky21,
       author = {{Lyubarsky}, Yuri},
        title = "{Emission Mechanisms of Fast Radio Bursts}",
      journal = {Universe},
         year = 2021,
        month = mar,
       volume = {7},
       number = {3},
          eid = {56},
        pages = {56},
          doi = {10.3390/universe7030056},
archivePrefix = {arXiv},
       eprint = {2103.00470},
 primaryClass = {astro-ph.HE},
       adsurl = {https://ui.adsabs.harvard.edu/abs/2021Univ....7...56L}
}

@ARTICLE{chime20,
       author = {{Andersen {\it et al.} (CHIME/FRB Collaboration)}, B.~C.},
        title = "{A bright millisecond-duration radio burst from a Galactic magnetar}",
      journal = {\nat},
         year = 2020,
        month = nov,
       volume = {587},
       number = {7832},
        pages = {54-58},
          doi = {10.1038/s41586-020-2863-y},
archivePrefix = {arXiv},
       eprint = {2005.10324},
 primaryClass = {astro-ph.HE},
       adsurl = {https://ui.adsabs.harvard.edu/abs/2020Natur.587...54C}
}

@ARTICLE{cordes16,
       author = {{Cordes}, J.~M. and {Wharton}, R.~S. and {Spitler}, L.~G. and {Chatterjee}, S. and {Wasserman}, I.},
        title = "{Radio Wave Propagation and the Provenance of Fast Radio Bursts}",
      journal = {arXiv e-prints},
         year = 2016,
        month = may,
          eid = {arXiv:1605.05890},
        pages = {arXiv:1605.05890},
          doi = {10.48550/arXiv.1605.05890},
archivePrefix = {arXiv},
       eprint = {1605.05890},
 primaryClass = {astro-ph.HE},
       adsurl = {https://ui.adsabs.harvard.edu/abs/2016arXiv160505890C}
}

@ARTICLE{chime21,
       author = {{Amiri {\it et al.} (CHIME/FRB Collaboration)}, Mandana},
        title = "{The First CHIME/FRB Fast Radio Burst Catalog}",
      journal = {\apjs},
         year = 2021,
        month = dec,
       volume = {257},
       number = {2},
          eid = {59},
        pages = {59},
          doi = {10.3847/1538-4365/ac33ab},
archivePrefix = {arXiv},
       eprint = {2106.04352},
 primaryClass = {astro-ph.HE},
       adsurl = {https://ui.adsabs.harvard.edu/abs/2021ApJS..257...59C}
}

@ARTICLE{kumar20a,
       author = {{Kumar}, Pawan and {Bo{\v{s}}njak}, {\v{Z}}eljka},
        title = "{FRB coherent emission from decay of Alfv{\'e}n waves}",
      journal = {\mnras},
         year = 2020,
        month = may,
       volume = {494},
       number = {2},
        pages = {2385-2395},
          doi = {10.1093/mnras/staa774},
archivePrefix = {arXiv},
       eprint = {2004.00644},
 primaryClass = {astro-ph.HE},
       adsurl = {https://ui.adsabs.harvard.edu/abs/2020MNRAS.494.2385K}
}

@ARTICLE{kumar17,
       author = {{Kumar}, Pawan and {Lu}, Wenbin and {Bhattacharya}, Mukul},
        title = "{Fast radio burst source properties and curvature radiation model}",
      journal = {\mnras},
         year = 2017,
        month = jul,
       volume = {468},
       number = {3},
        pages = {2726-2739},
          doi = {10.1093/mnras/stx665},
archivePrefix = {arXiv},
       eprint = {1703.06139},
 primaryClass = {astro-ph.HE},
       adsurl = {https://ui.adsabs.harvard.edu/abs/2017MNRAS.468.2726K}
}

@ARTICLE{kumar20b,
       author = {{Kumar}, Pawan and {Lu}, Wenbin},
        title = "{Radiation forces constrain the FRB mechanism}",
      journal = {\mnras},
         year = 2020,
        month = may,
       volume = {494},
       number = {1},
        pages = {1217-1228},
          doi = {10.1093/mnras/staa801},
archivePrefix = {arXiv},
       eprint = {2004.00645},
 primaryClass = {astro-ph.HE},
       adsurl = {https://ui.adsabs.harvard.edu/abs/2020MNRAS.494.1217K}
}

@ARTICLE{katz14,
       author = {{Katz}, J.~I.},
        title = "{Coherent emission in fast radio bursts}",
      journal = {\prd},
         year = 2014,
        month = may,
       volume = {89},
       number = {10},
          eid = {103009},
        pages = {103009},
          doi = {10.1103/PhysRevD.89.103009},
archivePrefix = {arXiv},
       eprint = {1309.3538},
 primaryClass = {astro-ph.HE},
}

@ARTICLE{lu18,
       author = {{Lu}, Wenbin and {Kumar}, Pawan},
        title = "{On the radiation mechanism of repeating fast radio bursts}",
      journal = {\mnras},
         year = 2018,
        month = jun,
       volume = {477},
       number = {2},
        pages = {2470-2493},
          doi = {10.1093/mnras/sty716},
archivePrefix = {arXiv},
       eprint = {1710.10270},
 primaryClass = {astro-ph.HE},
       adsurl = {https://ui.adsabs.harvard.edu/abs/2018MNRAS.477.2470L}
}

@ARTICLE{lu20,
       author = {{Lu}, Wenbin and {Kumar}, Pawan and {Zhang}, Bing},
        title = "{A unified picture of Galactic and cosmological fast radio bursts}",
      journal = {\mnras},
         year = 2020,
        month = oct,
       volume = {498},
       number = {1},
        pages = {1397-1405},
          doi = {10.1093/mnras/staa2450},
archivePrefix = {arXiv},
       eprint = {2005.06736},
 primaryClass = {astro-ph.HE},
       adsurl = {https://ui.adsabs.harvard.edu/abs/2020MNRAS.498.1397L}
}

@ARTICLE{snzhang20,
       author = {{Zhang {\it et al.}}, S. -N.},
        title = "{Insight-HXMT detection of a bright short x-ray counterpart of the Fast Radio Burst from SGR 1935+2154}",
      journal = {The Astronomer's Telegram},
         year = 2020,
        month = apr,
       volume = {13687},
        pages = {1},
       adsurl = {https://ui.adsabs.harvard.edu/abs/2020ATel13687....1Z}
}

@ARTICLE{yuan20,
       author = {{Yuan}, Yajie and {Beloborodov}, Andrei M. and {Chen}, Alexander Y. and {Levin}, Yuri},
        title = "{Plasmoid Ejection by Alfv{\'e}n Waves and the Fast Radio Bursts from SGR 1935+2154}",
      journal = {\apjl},
         year = 2020,
        month = sep,
       volume = {900},
       number = {2},
          eid = {L21},
        pages = {L21},
          doi = {10.3847/2041-8213/abafa8},
archivePrefix = {arXiv},
       eprint = {2006.04649},
 primaryClass = {astro-ph.HE},
       adsurl = {https://ui.adsabs.harvard.edu/abs/2020ApJ...900L..21Y}
}

@ARTICLE{qu24,
       author = {{Qu}, Yuanhong and {Zhang}, Bing},
        title = "{Coherent Inverse Compton Scattering in Fast Radio Bursts Revisited}",
      journal = {\apj},
         year = 2024,
        month = sep,
       volume = {972},
       number = {1},
          eid = {124},
        pages = {124},
          doi = {10.3847/1538-4357/ad5d5b},
archivePrefix = {arXiv},
       eprint = {2404.11948},
 primaryClass = {astro-ph.HE},
       adsurl = {https://ui.adsabs.harvard.edu/abs/2024ApJ...972..124Q}
}

@ARTICLE{bzhang17,
       author = {{Zhang}, Bing},
        title = "{A {\textquotedblleft}Cosmic Comb{\textquotedblright} Model of Fast Radio Bursts}",
      journal = {\apjl},
         year = 2017,
        month = feb,
       volume = {836},
       number = {2},
          eid = {L32},
        pages = {L32},
          doi = {10.3847/2041-8213/aa5ded},
archivePrefix = {arXiv},
       eprint = {1701.04094},
 primaryClass = {astro-ph.HE},
       adsurl = {https://ui.adsabs.harvard.edu/abs/2017ApJ...836L..32Z}
}

@ARTICLE{ioka20,
       author = {{Ioka}, Kunihito},
        title = "{Fast Radio Burst Breakouts from Magnetar Burst Fireballs}",
      journal = {\apjl},
         year = 2020,
        month = dec,
       volume = {904},
       number = {2},
          eid = {L15},
        pages = {L15},
          doi = {10.3847/2041-8213/abc6a3},
archivePrefix = {arXiv},
       eprint = {2008.01114},
 primaryClass = {astro-ph.HE},
       adsurl = {https://ui.adsabs.harvard.edu/abs/2020ApJ...904L..15I}
}

@ARTICLE{cooper21,
       author = {{Cooper}, A.~J. and {Wijers}, R.~A.~M.~J.},
        title = "{Coherent curvature radiation: maximum luminosity and high-energy emission}",
      journal = {\mnras},
         year = 2021,
        month = nov,
       volume = {508},
       number = {1},
        pages = {L32-L36},
          doi = {10.1093/mnrasl/slab099},
archivePrefix = {arXiv},
       eprint = {2108.07818},
 primaryClass = {astro-ph.HE},
       adsurl = {https://ui.adsabs.harvard.edu/abs/2021MNRAS.508L..32C}
}

@ARTICLE{lyubarsky14,
       author = {{Lyubarsky}, Yu.},
        title = "{A model for fast extragalactic radio bursts.}",
      journal = {\mnras},
         year = 2014,
        month = jul,
       volume = {442},
        pages = {L9-L13},
          doi = {10.1093/mnrasl/slu046},
archivePrefix = {arXiv},
       eprint = {1401.6674},
 primaryClass = {astro-ph.HE},
       adsurl = {https://ui.adsabs.harvard.edu/abs/2014MNRAS.442L...9L}
}

@ARTICLE{beloborodov20,
       author = {{Beloborodov}, Andrei M.},
        title = "{Blast Waves from Magnetar Flares and Fast Radio Bursts}",
      journal = {\apj},
         year = 2020,
        month = jun,
       volume = {896},
       number = {2},
          eid = {142},
        pages = {142},
          doi = {10.3847/1538-4357/ab83eb},
archivePrefix = {arXiv},
       eprint = {1908.07743},
 primaryClass = {astro-ph.HE},
       adsurl = {https://ui.adsabs.harvard.edu/abs/2020ApJ...896..142B}
}

@ARTICLE{ghisellini18,
       author = {{Ghisellini}, Gabriele and {Locatelli}, Nicola},
        title = "{Coherent curvature radiation and fast radio bursts}",
      journal = {\aap},
         year = 2018,
        month = jun,
       volume = {613},
          eid = {A61},
        pages = {A61},
          doi = {10.1051/0004-6361/201731820},
archivePrefix = {arXiv},
       eprint = {1708.07507},
 primaryClass = {astro-ph.HE},
       adsurl = {https://ui.adsabs.harvard.edu/abs/2018A&A...613A..61G}
}

@ARTICLE{beloborodov17,
       author = {{Beloborodov}, Andrei M.},
        title = "{A Flaring Magnetar in FRB 121102?}",
      journal = {\apjl},
         year = 2017,
        month = jul,
       volume = {843},
       number = {2},
          eid = {L26},
        pages = {L26},
          doi = {10.3847/2041-8213/aa78f3},
archivePrefix = {arXiv},
       eprint = {1702.08644},
 primaryClass = {astro-ph.HE},
       adsurl = {https://ui.adsabs.harvard.edu/abs/2017ApJ...843L..26B}
}

@ARTICLE{metzger19,
       author = {{Metzger}, Brian D. and {Margalit}, Ben and {Sironi}, Lorenzo},
        title = "{Fast radio bursts as synchrotron maser emission from decelerating relativistic blast waves}",
      journal = {\mnras},
         year = 2019,
        month = may,
       volume = {485},
       number = {3},
        pages = {4091-4106},
          doi = {10.1093/mnras/stz700},
archivePrefix = {arXiv},
       eprint = {1902.01866},
 primaryClass = {astro-ph.HE},
       adsurl = {https://ui.adsabs.harvard.edu/abs/2019MNRAS.485.4091M}
}

@ARTICLE{margalit20a,
       author = {{Margalit}, Ben and {Metzger}, Brian D. and {Sironi}, Lorenzo},
        title = "{Constraints on the engines of fast radio bursts}",
      journal = {\mnras},
         year = 2020,
        month = jun,
       volume = {494},
       number = {4},
        pages = {4627-4644},
          doi = {10.1093/mnras/staa1036},
archivePrefix = {arXiv},
       eprint = {1911.05765},
 primaryClass = {astro-ph.HE},
       adsurl = {https://ui.adsabs.harvard.edu/abs/2020MNRAS.494.4627M}
}

@ARTICLE{margalit20b,
       author = {{Margalit}, Ben and {Beniamini}, Paz and {Sridhar}, Navin and {Metzger}, Brian D.},
        title = "{Implications of a Fast Radio Burst from a Galactic Magnetar}",
      journal = {\apjl},
         year = 2020,
        month = aug,
       volume = {899},
       number = {2},
          eid = {L27},
        pages = {L27},
          doi = {10.3847/2041-8213/abac57},
archivePrefix = {arXiv},
       eprint = {2005.05283},
 primaryClass = {astro-ph.HE},
       adsurl = {https://ui.adsabs.harvard.edu/abs/2020ApJ...899L..27M}
}

@ARTICLE{waxman17,
       author = {{Waxman}, Eli},
        title = "{On the Origin of Fast Radio Bursts (FRBs)}",
      journal = {\apj},
         year = 2017,
        month = jun,
       volume = {842},
       number = {1},
          eid = {34},
        pages = {34},
          doi = {10.3847/1538-4357/aa713e},
archivePrefix = {arXiv},
       eprint = {1703.06723},
 primaryClass = {astro-ph.HE},
       adsurl = {https://ui.adsabs.harvard.edu/abs/2017ApJ...842...34W}
}

@ARTICLE{iwamoto24,
       author = {{Iwamoto}, Masanori and {Matsumoto}, Yosuke and {Amano}, Takanobu and {Matsukiyo}, Shuichi and {Hoshino}, Masahiro},
        title = "{Linearly Polarized Coherent Emission from Relativistic Magnetized Ion-Electron Shocks}",
      journal = {\prl},
         year = 2024,
        month = jan,
       volume = {132},
       number = {3},
          eid = {035201},
        pages = {035201},
          doi = {10.1103/PhysRevLett.132.035201},
archivePrefix = {arXiv},
       eprint = {2311.18487},
 primaryClass = {astro-ph.HE},
       adsurl = {https://ui.adsabs.harvard.edu/abs/2024PhRvL.132c5201I}
}

@ARTICLE{vanthieghem25a,
       author = {{Vanthieghem}, A. and {Levinson}, A.},
        title = "{Fast Radio Bursts as Precursor Radio Emission from Monster Shocks}",
      journal = {\prl},
         year = 2025,
        month = jan,
       volume = {134},
       number = {3},
          eid = {035201},
        pages = {035201},
          doi = {10.1103/PhysRevLett.134.035201},
archivePrefix = {arXiv},
       eprint = {2407.15076},
 primaryClass = {astro-ph.HE},
       adsurl = {https://ui.adsabs.harvard.edu/abs/2025PhRvL.134c5201V}
}

@ARTICLE{sobacchi24,
       author = {{Sobacchi}, E. and {Iwamoto}, M. and {Sironi}, L. and {Piran}, T.},
        title = "{Escape of fast radio bursts from magnetars}",
      journal = {\aap},
         year = 2024,
        month = oct,
       volume = {690},
          eid = {A332},
        pages = {A332},
          doi = {10.1051/0004-6361/202451725},
archivePrefix = {arXiv},
       eprint = {2409.10732},
 primaryClass = {astro-ph.HE},
       adsurl = {https://ui.adsabs.harvard.edu/abs/2024A&A...690A.332S}
}

@ARTICLE{nishiura24,
       author = {{Nishiura}, Rei and {Ioka}, Kunihito},
        title = "{Collective Thomson scattering in magnetized electron and positron pair plasma and the application to induced Compton scattering}",
      journal = {\prd},
         year = 2024,
        month = feb,
       volume = {109},
       number = {4},
          eid = {043048},
        pages = {043048},
          doi = {10.1103/PhysRevD.109.043048},
archivePrefix = {arXiv},
       eprint = {2310.02306},
 primaryClass = {astro-ph.HE},
       adsurl = {https://ui.adsabs.harvard.edu/abs/2024PhRvD.109d3048N}
}

@ARTICLE{matsukiyo09,
       author = {{Matsukiyo}, S. and {Hada}, T.},
        title = "{Relativistic Particle Acceleration in Developing Alfv{\'E}n Turbulence}",
      journal = {\apj},
         year = 2009,
        month = feb,
       volume = {692},
       number = {2},
        pages = {1004-1012},
          doi = {10.1088/0004-637X/692/2/1004},
archivePrefix = {arXiv},
       eprint = {0912.2154},
 primaryClass = {astro-ph.HE},
       adsurl = {https://ui.adsabs.harvard.edu/abs/2009ApJ...692.1004M}
}

@ARTICLE{lopez14,
       author = {{L{\'o}pez}, Rodrigo A. and {Mu{\~n}oz}, V{\'\i}ctor and {Vi{\~n}as}, Adolfo F. and {Alejandro Valdivia}, J.},
        title = "{Particle-in-cell simulation for parametric decays of a circularly polarized Alfv{\'e}n wave in relativistic thermal electron-positron plasma}",
      journal = {Physics of Plasmas},
         year = 2014,
        month = mar,
       volume = {21},
       number = {3},
          eid = {032102},
        pages = {032102},
          doi = {10.1063/1.4867255},
       adsurl = {https://ui.adsabs.harvard.edu/abs/2014PhPl...21c2102L}
}

@ARTICLE{munoz14,
       author = {{Mu{\~n}oz}, V. and {Asenjo}, F.~A. and {Dom{\'\i}nguez}, M. and {L{\'o}pez}, R.~A. and {Valdivia}, J.~A. and {Vi{\~n}as}, A. and {Hada}, T.},
        title = "{Large-amplitude electromagnetic waves in magnetized relativistic plasmas with temperature}",
      journal = {Nonlinear Processes in Geophysics},
         year = 2014,
        month = feb,
       volume = {21},
       number = {1},
        pages = {217-236},
          doi = {10.5194/npg-21-217-2014},
       adsurl = {https://ui.adsabs.harvard.edu/abs/2014NPGeo..21..217M}
}

@Article{numpy,
 title = {Array programming with {NumPy}},
 author = {Charles R. Harris and K. Jarrod Millman and St{'{e}}fan J. van der Walt and Ralf Gommers and Pauli Virtanen and David
                 Cournapeau and Eric Wieser and Julian Taylor and Sebastian
                 Berg and Nathaniel J. Smith and Robert Kern and Matti Picus
                 and Stephan Hoyer and Marten H. van Kerkwijk and Matthew
                 Brett and Allan Haldane and Jaime Fern{'{a}}ndez del
                 R{'{\i}}o and Mark Wiebe and Pearu Peterson and Pierre
                 G{'{e}}rard-Marchant and Kevin Sheppard and Tyler Reddy and
                 Warren Weckesser and Hameer Abbasi and Christoph Gohlke and
                 Travis E. Oliphant},
 year          = {2020},
 month         = sep,
 journal       = {Nature},
 volume        = {585},
 number        = {7825},
 pages         = {357--362},
 doi           = {10.1038/s41586-020-2649-2},
 publisher     = {Springer Science and Business Media {LLC}},
 url           = {https://doi.org/10.1038/s41586-020-2649-2}
}

@ARTICLE{lee83,
       author = {{Lee}, N.~C. and {Parks}, G.~K.},
        title = "{Ponderomotive force in a warm two-fluid plasma}",
      journal = {Physics of Fluids},
         year = 1983,
        month = mar,
       volume = {26},
       number = {3},
        pages = {724-729},
          doi = {10.1063/1.864196},
       adsurl = {https://ui.adsabs.harvard.edu/abs/1983PhFl...26..724L}
}

@ARTICLE{lee96,
       author = {{Lee}, Nam C. and {Parks}, George K.},
        title = "{Ponderomotive acceleration of ions by circularly polarized electromagnetic waves}",
      journal = {\grl},
         year = 1996,
        month = feb,
       volume = {23},
       number = {4},
        pages = {327-330},
          doi = {10.1029/96GL00157},
       adsurl = {https://ui.adsabs.harvard.edu/abs/1996GeoRL..23..327L}
}

@ARTICLE{klima66,
       author = {{Kl{\'\i}ma}, R.},
        title = "{{\CYRO} {\CYRD}{\CYRV}{\CYRI}{\CYRZH}{\CYRE}{\CYRN}{\CYRI}{\CYRI} {\CYRCH}{\CYRA}{\CYRS}{\CYRT}{\CYRI}{\CYRC} {\CYRV} {\CYRN}{\CYRE}{\CYRR}{\CYRE}{\CYRZ}{\CYRO}{\CYRN}{\CYRA}{\CYRN}{\CYRS}{\CYRN}{\CYRERY}{\CYRH} {\CYRV}. {\CYRCH}. {\CYRI} {\CYRM}{\CYRA}{\CYRG}{\CYRN}{\CYRI}{\CYRT}{\CYRO}{\CYRS}{\CYRT}{\CYRA}{\CYRT}{\CYRI}{\CYRCH}{\CYRE}{\CYRS}{\CYRK}{\CYRI}{\CYRH} {\CYRP}{\CYRO}{\CYRL}{\CYRYA}{\CYRH}}",
      journal = {Czechoslovak Journal of Physics},
         year = 1966,
        month = aug,
       volume = {16},
       number = {8},
        pages = {681-696},
          doi = {10.1007/BF01689569},
       adsurl = {https://ui.adsabs.harvard.edu/abs/1966CzJPh..16..681.}
}

@ARTICLE{klima68,
       author = {{Kl{\'\i}ma}, R.},
        title = "{The drifts and hydrodynamics of particles in a field with a high-frequency component}",
      journal = {Czechoslovak Journal of Physics},
         year = 1968,
        month = oct,
       volume = {18},
       number = {10},
        pages = {1280-1291},
          doi = {10.1007/BF01690802},
       adsurl = {https://ui.adsabs.harvard.edu/abs/1968CzJPh..18.1280K}
}

@ARTICLE{hatori81,
       author = {{Hatori}, T. and {Washimi}, H.},
        title = "{Covariant Form of the Ponderomotive Potentials in a Magnetized Plasma}",
      journal = {\prl},
         year = 1981,
        month = jan,
       volume = {46},
       number = {4},
        pages = {240-243},
          doi = {10.1103/PhysRevLett.46.240},
       adsurl = {https://ui.adsabs.harvard.edu/abs/1981PhRvL..46..240H}
}

@ARTICLE{cary77,
       author = {{Cary}, J.~R. and {Kaufman}, A.~N.},
        title = "{Ponderomotive Force and Linear Susceptibility in Vlasov Plasma}",
      journal = {\prl},
         year = 1977,
        month = aug,
       volume = {39},
       number = {7},
        pages = {402-404},
          doi = {10.1103/PhysRevLett.39.402},
       adsurl = {https://ui.adsabs.harvard.edu/abs/1977PhRvL..39..402C}
}

@ARTICLE{cary81,
       author = {{Cary}, J.~R. and {Kaufman}, A.~N.},
        title = "{Ponderomotive effects in collisionless plasma: A Lie transform approach}",
      journal = {Physics of Fluids},
         year = 1981,
        month = jul,
       volume = {24},
       number = {7},
        pages = {1238-1250},
          doi = {10.1063/1.863527},
       adsurl = {https://ui.adsabs.harvard.edu/abs/1981PhFl...24.1238C}
}

@ARTICLE{tabak94,
       author = {{Tabak}, Max and {Hammer}, James and {Glinsky}, Michael E. and {Kruer}, William L. and {Wilks}, Scott C. and {Woodworth}, John and {Campbell}, E. Michael and {Perry}, Michael D. and {Mason}, Rodney J.},
        title = "{Ignition and high gain with ultrapowerful lasers*}",
      journal = {Physics of Plasmas},
         year = 1994,
        month = may,
       volume = {1},
       number = {5},
        pages = {1626-1634},
          doi = {10.1063/1.870664},
       adsurl = {https://ui.adsabs.harvard.edu/abs/1994PhPl....1.1626T}
}

@ARTICLE{deutsch96,
       author = {{Deutsch}, C. and {Furukawa}, H. and {Mima}, K. and {Murakami}, M. and {Nishihara}, K.},
        title = "{Interaction Physics of the Fast Ignitor Concept}",
      journal = {\prl},
         year = 1996,
        month = sep,
       volume = {77},
       number = {12},
        pages = {2483-2486},
          doi = {10.1103/PhysRevLett.77.2483},
       adsurl = {https://ui.adsabs.harvard.edu/abs/1996PhRvL..77.2483D}
}

@ARTICLE{kwan79,
       author = {{Kwan}, T. and {Dawson}, J.~M.},
        title = "{Investigation of the free electron laser with a guide magnetic field}",
      journal = {Physics of Fluids},
         year = 1979,
        month = jun,
       volume = {22},
       number = {6},
        pages = {1089-1103},
          doi = {10.1063/1.862702},
       adsurl = {https://ui.adsabs.harvard.edu/abs/1979PhFl...22.1089K}
}

@ARTICLE{friedland80,
       author = {{Friedland}, L.},
        title = "{Electron beam dynamics in combined guide and pump magnetic fields for free electron laser applications}",
      journal = {Physics of Fluids},
         year = 1980,
        month = dec,
       volume = {23},
       number = {12},
        pages = {2376-2382},
          doi = {10.1063/1.862942},
       adsurl = {https://ui.adsabs.harvard.edu/abs/1980PhFl...23.2376F}
}

@ARTICLE{thornton13,
       author = {{Thornton}, D. and {Stappers}, B. and {Bailes}, M. and {Barsdell}, B. and {Bates}, S. and {Bhat}, N.~D.~R. and {Burgay}, M. and {Burke-Spolaor}, S. and {Champion}, D.~J. and {Coster}, P. and {D'Amico}, N. and {Jameson}, A. and {Johnston}, S. and {Keith}, M. and {Kramer}, M. and {Levin}, L. and {Milia}, S. and {Ng}, C. and {Possenti}, A. and {van Straten}, W.},
        title = "{A Population of Fast Radio Bursts at Cosmological Distances}",
      journal = {Science},
         year = 2013,
        month = jul,
       volume = {341},
       number = {6141},
        pages = {53-56},
          doi = {10.1126/science.1236789},
archivePrefix = {arXiv},
       eprint = {1307.1628},
 primaryClass = {astro-ph.HE},
       adsurl = {https://ui.adsabs.harvard.edu/abs/2013Sci...341...53T}
}

@ARTICLE{mereghetti20,
       author = {{Mereghetti}, S. and {Savchenko}, V. and {Ferrigno}, C. and {G{\"o}tz}, D. and {Rigoselli}, M. and {Tiengo}, A. and {Bazzano}, A. and {Bozzo}, E. and {Coleiro}, A. and {Courvoisier}, T.~J.-L. and {Doyle}, M. and {Goldwurm}, A. and {Hanlon}, L. and {Jourdain}, E. and {von Kienlin}, A. and {Lutovinov}, A. and {Martin-Carrillo}, A. and {Molkov}, S. and {Natalucci}, L. and {Onori}, F. and {Panessa}, F. and {Rodi}, J. and {Rodriguez}, J. and {S{\'a}nchez-Fern{\'a}ndez}, C. and {Sunyaev}, R. and {Ubertini}, P.},
        title = "{INTEGRAL Discovery of a Burst with Associated Radio Emission from the Magnetar SGR 1935+2154}",
      journal = {\apjl},
         year = 2020,
        month = aug,
       volume = {898},
       number = {2},
          eid = {L29},
        pages = {L29},
          doi = {10.3847/2041-8213/aba2cf},
archivePrefix = {arXiv},
       eprint = {2005.06335},
 primaryClass = {astro-ph.HE},
       adsurl = {https://ui.adsabs.harvard.edu/abs/2020ApJ...898L..29M}
}

@ARTICLE{bochenek20,
       author = {{Bochenek}, C.~D. and {Ravi}, V. and {Belov}, K.~V. and {Hallinan}, G. and {Kocz}, J. and {Kulkarni}, S.~R. and {McKenna}, D.~L.},
        title = "{A fast radio burst associated with a Galactic magnetar}",
      journal = {\nat},
         year = 2020,
        month = nov,
       volume = {587},
       number = {7832},
        pages = {59-62},
          doi = {10.1038/s41586-020-2872-x},
archivePrefix = {arXiv},
       eprint = {2005.10828},
 primaryClass = {astro-ph.HE},
       adsurl = {https://ui.adsabs.harvard.edu/abs/2020Natur.587...59B}
}

@ARTICLE{li21,
       author = {{Li}, C.~K. and {Lin}, L. and {Xiong}, S.~L. and {Ge}, M.~Y. and {Li}, X.~B. and {Li}, T.~P. and {Lu}, F.~J. and {Zhang}, S.~N. and {Tuo}, Y.~L. and {Nang}, Y. and {Zhang}, B. and {Xiao}, S. and {Chen}, Y. and {Song}, L.~M. and {Xu}, Y.~P. and {Liu}, C.~Z. and {Jia}, S.~M. and {Cao}, X.~L. and {Qu}, J.~L. and {Zhang}, S. and {Gu}, Y.~D. and {Liao}, J.~Y. and {Zhao}, X.~F. and {Tan}, Y. and {Nie}, J.~Y. and {Zhao}, H.~S. and {Zheng}, S.~J. and {Zheng}, Y.~G. and {Luo}, Q. and {Cai}, C. and {Li}, B. and {Xue}, W.~C. and {Bu}, Q.~C. and {Chang}, Z. and {Chen}, G. and {Chen}, L. and {Chen}, T.~X. and {Chen}, Y.~B. and {Chen}, Y.~P. and {Cui}, W. and {Cui}, W.~W. and {Deng}, J.~K. and {Dong}, Y.~W. and {Du}, Y.~Y. and {Fu}, M.~X. and {Gao}, G.~H. and {Gao}, H. and {Gao}, M. and {Gu}, Y.~D. and {Guan}, J. and {Guo}, C.~C. and {Han}, D.~W. and {Huang}, Y. and {Huo}, J. and {Jiang}, L.~H. and {Jiang}, W.~C. and {Jin}, J. and {Jin}, Y.~J. and {Kong}, L.~D. and {Li}, G. and {Li}, M.~S. and {Li}, W. and {Li}, X. and {Li}, X.~F. and {Li}, Y.~G. and {Li}, Z.~W. and {Liang}, X.~H. and {Liu}, B.~S. and {Liu}, G.~Q. and {Liu}, H.~W. and {Liu}, X.~J. and {Liu}, Y.~N. and {Lu}, B. and {Lu}, X.~F. and {Luo}, T. and {Ma}, X. and {Meng}, B. and {Ou}, G. and {Sai}, N. and {Shang}, R.~C. and {Song}, X.~Y. and {Sun}, L. and {Tao}, L. and {Wang}, C. and {Wang}, G.~F. and {Wang}, J. and {Wang}, W.~S. and {Wang}, Y.~S. and {Wen}, X.~Y. and {Wu}, B.~B. and {Wu}, B.~Y. and {Wu}, M. and {Xiao}, G.~C. and {Xu}, H. and {Yang}, J.~W. and {Yang}, S. and {Yang}, Y.~J. and {Yang}, Yi-Jung and {Yi}, Q.~B. and {Yin}, Q.~Q. and {You}, Y. and {Zhang}, A.~M. and {Zhang}, C.~M. and {Zhang}, F. and {Zhang}, H.~M. and {Zhang}, J. and {Zhang}, T. and {Zhang}, W. and {Zhang}, W.~C. and {Zhang}, W.~Z. and {Zhang}, Y. and {Zhang}, Yue and {Zhang}, Y.~F. and {Zhang}, Y.~J. and {Zhang}, Z. and {Zhang}, Zhi and {Zhang}, Z.~L. and {Zhou}, D.~K. and {Zhou}, J.~F. and {Zhu}, Y. and {Zhu}, Y.~X. and {Zhuang}, R.~L.},
        title = "{HXMT identification of a non-thermal X-ray burst from SGR J1935+2154 and with FRB 200428}",
      journal = {Nature Astronomy},
         year = 2021,
        month = apr,
       volume = {5},
        pages = {378-384},
          doi = {10.1038/s41550-021-01302-6},
archivePrefix = {arXiv},
       eprint = {2005.11071},
 primaryClass = {astro-ph.HE},
       adsurl = {https://ui.adsabs.harvard.edu/abs/2021NatAs...5..378L}
}

@ARTICLE{ridnaia21,
       author = {{Ridnaia}, A. and {Svinkin}, D. and {Frederiks}, D. and {Bykov}, A. and {Popov}, S. and {Aptekar}, R. and {Golenetskii}, S. and {Lysenko}, A. and {Tsvetkova}, A. and {Ulanov}, M. and {Cline}, T.~L.},
        title = "{A peculiar hard X-ray counterpart of a Galactic fast radio burst}",
      journal = {Nature Astronomy},
         year = 2021,
        month = apr,
       volume = {5},
        pages = {372-377},
          doi = {10.1038/s41550-020-01265-0},
archivePrefix = {arXiv},
       eprint = {2005.11178},
 primaryClass = {astro-ph.HE},
       adsurl = {https://ui.adsabs.harvard.edu/abs/2021NatAs...5..372R}
}

@ARTICLE{thompson98,
       author = {{Thompson}, Christopher and {Blaes}, Omer},
        title = "{Magnetohydrodynamics in the extreme relativistic limit}",
      journal = {\prd},
         year = 1998,
        month = mar,
       volume = {57},
       number = {6},
        pages = {3219-3234},
          doi = {10.1103/PhysRevD.57.3219},
       adsurl = {https://ui.adsabs.harvard.edu/abs/1998PhRvD..57.3219T}
}

@ARTICLE{melrose17,
       author = {{Melrose}, D.~B.},
        title = "{Coherent emission mechanisms in astrophysical plasmas}",
      journal = {Reviews of Modern Plasma Physics},
         year = 2017,
        month = dec,
       volume = {1},
       number = {1},
          eid = {5},
        pages = {5},
          doi = {10.1007/s41614-017-0007-0},
archivePrefix = {arXiv},
       eprint = {1707.02009},
 primaryClass = {physics.plasm-ph},
       adsurl = {https://ui.adsabs.harvard.edu/abs/2017RvMPP...1....5M}
}

@ARTICLE{chen22,
       author = {{Chen}, Alexander Y. and {Yuan}, Yajie and {Li}, Xinyu and {Mahlmann}, Jens F.},
        title = "{Propagation of a Strong Fast Magnetosonic Wave in the Magnetosphere of a Neutron Star}",
      journal = {arXiv e-prints},
         year = 2022,
        month = oct,
          eid = {arXiv:2210.13506},
        pages = {arXiv:2210.13506},
          doi = {10.48550/arXiv.2210.13506},
archivePrefix = {arXiv},
       eprint = {2210.13506},
 primaryClass = {astro-ph.HE},
       adsurl = {https://ui.adsabs.harvard.edu/abs/2022arXiv221013506C}
}

@ARTICLE{beloborodov23,
       author = {{Beloborodov}, Andrei M.},
        title = "{Monster Radiative Shocks in the Perturbed Magnetospheres of Neutron Stars}",
      journal = {\apj},
         year = 2023,
        month = dec,
       volume = {959},
       number = {1},
          eid = {34},
        pages = {34},
          doi = {10.3847/1538-4357/acf659},
archivePrefix = {arXiv},
       eprint = {2210.13509},
 primaryClass = {astro-ph.HE},
       adsurl = {https://ui.adsabs.harvard.edu/abs/2023ApJ...959...34B}
}

@ARTICLE{beloborodov24,
       author = {{Beloborodov}, Andrei M.},
        title = "{Damping of Strong GHz Waves near Magnetars and the Origin of Fast Radio Bursts}",
      journal = {\apj},
         year = 2024,
        month = nov,
       volume = {975},
       number = {2},
          eid = {223},
        pages = {223},
          doi = {10.3847/1538-4357/ad698c},
archivePrefix = {arXiv},
       eprint = {2307.12182},
 primaryClass = {astro-ph.HE},
       adsurl = {https://ui.adsabs.harvard.edu/abs/2024ApJ...975..223B}
}

@ARTICLE{vanthieghem25,
       author = {{Vanthieghem}, A. and {Levinson}, A.},
        title = "{Fast Radio Bursts as Precursor Radio Emission from Monster Shocks}",
      journal = {\prl},
         year = 2025,
        month = jan,
       volume = {134},
       number = {3},
          eid = {035201},
        pages = {035201},
          doi = {10.1103/PhysRevLett.134.035201},
archivePrefix = {arXiv},
       eprint = {2407.15076},
 primaryClass = {astro-ph.HE},
       adsurl = {https://ui.adsabs.harvard.edu/abs/2025PhRvL.134c5201V}
}

\end{document}